\documentclass[aps,prl,twocolumn,superscriptaddress,floatfix,showpacs]{revtex4-2}

\usepackage{amsmath,amssymb}
\usepackage{graphicx}
\usepackage{bm}
\usepackage[hidelinks]{hyperref}
\usepackage{xcolor}

\begin{document}

\title{Free-fermion spectral structure enables strange nonchaotic attractor classification\\
in Aubry-Andr\'{e}-Harper quantum reservoirs}

\author{Suresh Kumarasamy}
\affiliation{Centre for Artificial Intelligence, Easwari Engineering College, Chennai 600089, Tamil Nadu, India}
\affiliation{Center for Cognitive Science, Trichy SRM Medical College Hospital and Research Center, Trichy, India}

\author{Dianavinnarasi Joseph}
\affiliation{Centre for Computational Biology, Easwari Engineering College, Chennai 600089, Tamil Nadu, India}

\author{Manish Dev Shrimali}
\affiliation{Department of Physics, Central University of Rajasthan, Ajmer 305 817, Rajasthan, India}

\author{Awadhesh Prasad}
\affiliation{Department of Physics and Astrophysics, University of Delhi, Delhi 110 007, India}
\date{\today}

\begin{abstract}
An Aubry-Andr\'{e}-Harper (AAH) quantum reservoir classifies strange nonchaotic attractors (SNAs) against chaos, substantially outperforming reservoirs built from dense random matrices on the full Fock space. Ablations and controlled interventions identify \emph{free-fermion one-body transition sparsity}, rather than proximity to a quantum phase transition, as the dominant factor under our controls, since $H_0$ is quadratic in fermion operators and only $N(N-1)2^{N-2}=3584$ of the $65{,}280$ Fock-space transitions carry weight. Random on-site disorder, which preserves this quadratic structure, matches AAH accuracy, whereas full-Hilbert-space random matrices and eigenvector permutations, which break it, collapse to the random baseline. We characterize the resulting frequency-selective filtering with a Hamiltonian transition kernel, validate it to $N=8$ qubits, and reproduce it in a second dynamical system. We also use the reservoir as a training-free screen for strange-nonchaotic structure in Kepler RR Lyrae light curves.
\end{abstract}


\maketitle

Quantum reservoir computing (QRC) uses the intrinsic dynamics of a quantum system as a high-dimensional nonlinear feature map~\cite{Fujii2017,Nakajima2019,Mujal2021}. An input signal drives the reservoir, and a linear readout, typically optimized by ridge regression, extracts target functions from the measured expectation values. Because the quantum evolution performs the nonlinear mapping, backpropagation through the reservoir is unnecessary, making QRC more efficient to train than deep quantum neural networks~\cite{Beer2020}.

A quantum reservoir's power derives from the exponential Hilbert space, where an $N$-qubit reservoir supplies up to $4^N$ observables~\cite{Ghosh2019,Nokkala2021}, but which Hamiltonian to choose remains open. Reservoirs near quantum phase transitions perform well~\cite{Martinez2021,Xia2022}, attributed to critical sensitivity, yet this picture does not identify \emph{which} spectral features set performance. We show that the dominant determinant for spectrally structured tasks is instead \emph{free-fermion one-body transition sparsity}, rather than proximity to a phase transition. Structured classical reservoirs (delay lines) show analogous frequency-selective advantages~\cite{Appeltant2011,Larger2012}, but no such spectral design principle has been established for quantum reservoirs.

Strange nonchaotic attractors (SNAs) arise in quasiperiodically forced dissipative systems and combine a fractal, non-differentiable geometry with nonpositive Lyapunov exponents~\cite{Grebogi1984,Feudel2006,Prasad2001,Romeiras1987}. They are therefore hard to separate from chaos by standard dynamical invariants, which overlap substantially between the two~\cite{Pikovsky1995}. Their power spectra are singular-continuous, with quasiperiodic peaks at combinations $n\omega_1+m\omega_2$ over a fractal background~\cite{Prasad2001}, whereas chaotic spectra are broadband and absolutely continuous, a spectral distinction we exploit for discrimination.

We study SNA classification with an Aubry-Andr\'{e}-Harper (AAH) quantum reservoir~\cite{Aubry1980,Harper1955}, a standard Hamiltonian for quasiperiodic tight-binding chains, with the incommensurability parameter set to the golden ratio $\varphi=(\sqrt{5}-1)/2$. The SNA-generating drive similarly employs this ratio. Ablation and controlled-intervention experiments identify this transition sparsity as the dominant observed factor driving performance. At matched conditions ($\Delta=2J$, $N=8$), AAH outperforms dense full-Hilbert-space random Hamiltonians by $\sim\!8$ percentage points ($89.8\%$ vs.\ $82.3\%$). A random quadratic (on-site disorder) Hamiltonian, which preserves both particle-number conservation and the one-body Slater-determinant selection rule, matches AAH accuracy ($89.9\%$ vs.\ $89.8\%$, the deterministic AAH value lying within the disorder ensemble confidence interval), whereas within-spectrum eigenvector permutations, which break this structure, collapse to the random-matrix baseline. We develop a Hamiltonian transition kernel and an associated nonlinear power spectral density (PSD) formalism to characterize the resulting frequency-selective filtering. Because the two reservoirs that retain the free-fermion structure also retain $U(1)$ particle-number conservation, our design does not isolate spectral sparsity \emph{within} the free-fermion class from particle conservation alone. We return to this caveat in the Discussion.

The AAH Hamiltonian~\cite{Aubry1980,Harper1955} on a lattice of $N$ sites (equivalently $N$ qubits) reads
\begin{equation}\label{eq:AAH}
  H_0 = -J\!\sum_{j=1}^{N-1}\!(c_j^\dagger c_{j+1}+\text{h.c.})
       + \Delta\!\sum_{j=1}^{N}\!\cos(2\pi\beta j+\phi_0)\,\hat{n}_j\,,
\end{equation}
where $J$ is the hopping amplitude, and $\Delta$ represents the strength of the quasiperiodic potential. Here, $\beta=\varphi=(\sqrt{5}-1)/2$ is the inverse golden mean, and $\phi_0$ is the initial phase, which is set to zero in the current work. The ratio $\Delta/J$ controls the localization transition, with delocalized eigenstates for $\Delta<2J$ and localized eigenstates for $\Delta>2J$. For irrational $\beta$ and any nonzero $\Delta$ the spectrum is a Cantor set~\cite{Avila2009}. At the self-dual critical point $\Delta=2J$ it also has zero Lebesgue measure and is multifractal. The integrated density of states (IDOS) within the spectral gaps takes values from the module $\mathbb{Z}+\mathbb{Z}\beta$, in accordance with the gap-labeling theorem~\cite{Bellissard1982}. These are dimensionless gap labels rather than transition frequencies.

The input signal $s(t)$ is coupled to the first site of the reservoir with coupling strength $\kappa$ via the term $H_\text{in}(t)=\kappa\,s(t)\,\hat{n}_1$. The evolution of the density matrix is governed by the Lindblad master equation~\cite{Lindblad1976,GKS1976}:
\begin{equation}\label{eq:lindblad}
  \dot{\rho} = -i[H_0+H_\text{in},\rho]
             + \sum_k \gamma_k\!\left(L_k\rho L_k^\dagger
               -\tfrac{1}{2}\{L_k^\dagger L_k,\rho\}\right),
\end{equation}
where the index $k$ labels the Lindblad channels, $\gamma_k$ their rates, and $L_k$ the corresponding jump operators, here weak dephasing and damping. Varying both rates by $\pm50\%$ changes accuracy by less than $3$ percentage points (SM~\cite{suppmat}, Fig.~\ref{fig:robustness}). For $N=8$ qubits the full $2^N$-dimensional Fock space yields $65{,}280$ possible transitions. Site occupancies $\langle\hat{n}_j\rangle$ and nearest-neighbor correlators $\langle\hat{n}_j\hat{n}_{j+1}\rangle$ ($15$ base observables for $N=8$) are recorded at each of $M=5$ virtual nodes per input step (time multiplexing) and pooled as mean and standard deviation over the processing window of $T$ input steps, giving $15\times2\times M=150$ observables fed to a ridge regression readout.

The discrimination signals are generated by the quasiperiodically forced Duffing
oscillator~\cite{Romeiras1987,Prasad2001} $\ddot{x}+\delta\dot{x}+\alpha x+\beta_D x^3
  = A_1\cos\omega_1 t + A_2\cos\omega_2 t$,
with $\delta=0.2$, $\alpha=-1$, $\beta_D=1$, $\omega_1=1$, $A_1=0.3$, and
$\omega_2/\omega_1=\varphi$.  Varying the forcing amplitude $A_2$ interpolates between SNA
($A_2\in[0.42,0.60]$, maximal Lyapunov exponent $\lambda_\text{max}<0$) and chaos
($A_2\in[0.68,0.90]$, $\lambda_\text{max}>0$).  We generate 200 signals (100
per class, length 1500) and average over 10 random 70/30 train--test splits.

\begin{figure}[tb]
  \centering
  \includegraphics[width=\columnwidth]{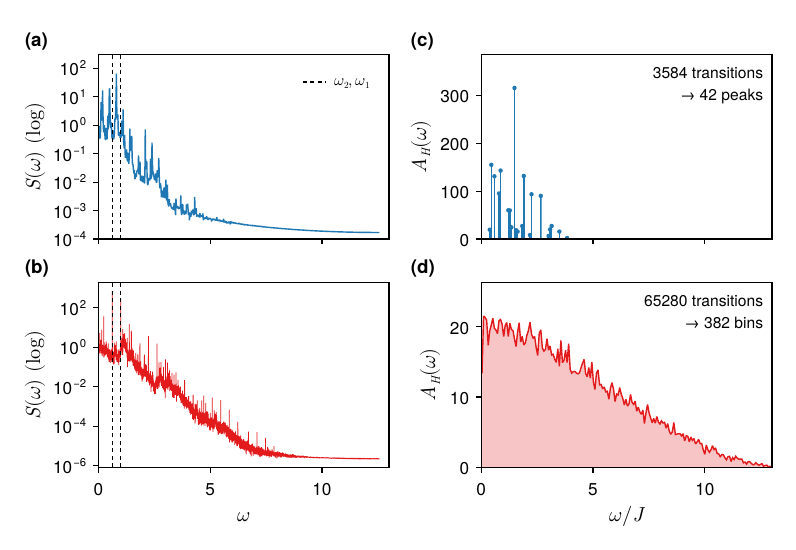}
  \caption{Signal spectra versus reservoir transition spectra ($J\!=\!\hbar\!=\!1$).
  (a)~Duffing PSD averaged over the SNA class ($A_2\in[0.42,0.60]$,
  $\lambda_\text{max}<0$); (b)~chaotic Duffing PSD at $A_2=0.78$
  ($\lambda_\text{max}>0$; ten trajectories, $2^{14}$ samples each).
  Dashed lines mark the drive frequencies $\omega_1\!=\!1$,
  $\omega_2\!=\!\varphi\,\omega_1$; signal power lies below
  $\omega\!\approx\!5$.
  (c,d)~Transition kernel $A_H(\omega)$ [Eq.~\eqref{eq:AH}] for the AAH
  delocalized phase ($\Delta=0.5J$) and a matched random (GUE) reservoir
  ($N\!=\!8$, $V=\kappa\hat{n}_1$): the 3584 active AAH transitions collapse
  onto $\sim\!42$ sharp peaks, whereas the GUE forms a quasi-continuum over
  all $65{,}280$. Phase portraits and full $A_H(\omega)$ for all four
  reservoir types are in the Supplemental Material~\cite{suppmat}.}
  \label{fig:fig1}
\end{figure}

Though their power spectra differ sharply [Fig.~\ref{fig:fig1}(a,b)], SNA and chaotic signals have similar $(x,\dot{x})$ phase portraits, and conventional chaos indicators (correlation dimension, recurrence statistics) are ambiguous near the SNA--chaos boundary (SM~\cite{suppmat}), making this a stringent test. The reservoir-side mechanism rests on the contrast in Figs.~\ref{fig:fig1}(c,d). The AAH transition kernel $A_H(\omega)$ collapses to a sparse comb of peaks while the GUE reservoir activates a dense quasi-continuum, a parallel between sparse signal and sparse reservoir spectra that anticipates the resonance mechanism below.

\begin{figure}[tb]
  \centering
  \includegraphics[width=\columnwidth]{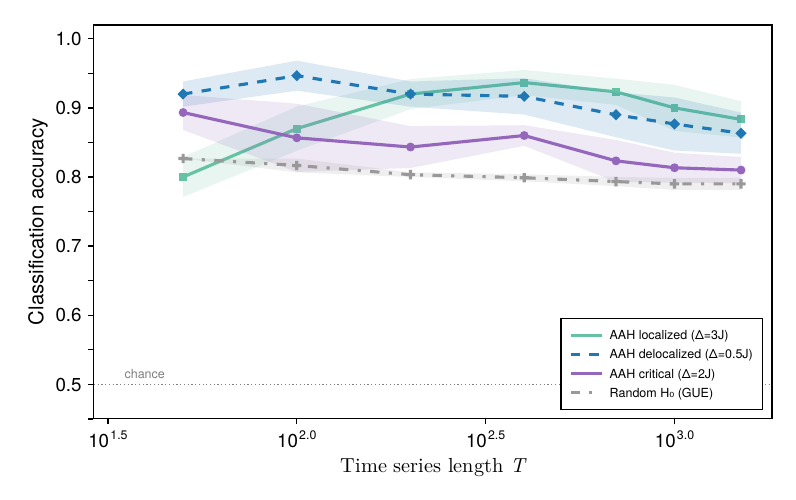}
  \caption{Discrimination accuracy versus processing horizon $T$ for the
  three AAH phases and a matched random (GUE) baseline.  Shaded bands:
  $\pm1\sigma$ over 10 random 70/30 train--test splits. Best-$T$ accuracies
  and the large-$T$ decline are discussed in the main text.}
  \label{fig:accuracy}
\end{figure}

The discrimination accuracy as a function of the processing horizon $T$ for the three AAH phases, along with a random-Hamiltonian baseline (GUE, with matched dimension and dissipation), is shown in Fig.~\ref{fig:accuracy}. All three AAH phases outperform the unstructured random reservoir: the delocalized phase ($\Delta=0.5J$) achieves an accuracy of $94.7\%$ at $T=100$, while the localized phase ($\Delta=3J$) reaches $93.7\%$ at $T=400$, and the critical phase ($\Delta=2J$) yields $89.8\%$ at $T=50$. In contrast, the random-Hamiltonian baseline saturates at $79.0\%$ on the broad-$T$ accuracy curve of Fig.~\ref{fig:accuracy}. The higher GUE numbers reported in the ablation ($83.5\%$) and controlled-intervention ($82.3\%$) studies below correspond to the best-$T$ accuracy computed on the matched-control protocols ($\Delta=2J$, $n=6$ seeds) and are not directly comparable to the broad-$T$ saturation here.

The approximately $16$-percentage-point gap between the AAH and random Hamiltonians shows that Hamiltonian \emph{structure}, beyond Hilbert-space dimension and dissipative dynamics, affects QRC performance on this task. The delocalized phase outperforms the critical phase despite the latter's more complex multifractal spectrum, which we analyze with linear response theory below. The $T$-dependence of accuracy reflects a mean-std pooling trade-off (SM~\cite{suppmat}, Fig.~\ref{fig:fisher_vs_T}). The standard linear memory capacity ($C\approx1.2$; SM Fig.~\ref{fig:memory_fisher}) is unreliable here because the driving is strongly nonlinear ($\kappa=2.0J$).

To analyze the underlying physics, we define the transition-frequency spectrum of the reservoir, weighted by the input coupling operator $V=\kappa\hat{n}_1$:
\begin{equation}\label{eq:AH}
  A_H(\omega) = \sum_{m\neq n}|\langle m|V|n\rangle|^2\,
                \delta(\omega-\omega_{mn})\,,
\end{equation}
where $\omega_{mn}=(E_m-E_n)/\hbar$ denotes the transition frequencies between the eigenstates $|m\rangle$ and $|n\rangle$ of the unperturbed Hamiltonian $H_0$. Equation~\eqref{eq:AH} is the unweighted transition density (a Fermi golden-rule kernel without thermal occupation factors or principal-value structure); we use it throughout as a structural descriptor of the closed-system frequency-selective response, distinct from the full Liouvillian susceptibility of the driven--dissipative problem. For the AAH phases, only 3584 out of the 65,280 possible transitions carry non-negligible weight, resulting in a highly structured and sparse spectrum [Fig.~\ref{fig:fig1}(c); full $A_H(\omega)$ for all four reservoirs in SM~\cite{suppmat}, Fig.~\ref{fig:transition_spectrum}]. This sparsity is exact and generic to non-interacting reservoirs. Because $H_0$ is quadratic in fermion operators its eigenstates are Slater determinants, and the one-body coupling $V=\kappa\hat{n}_1$ connects only those determinants differing by a single occupied orbital, so that summing the resulting single-excitation pairs over all particle-number sectors yields exactly $N(N-1)2^{N-2}=56\times 64=3584$ allowed transitions, which collapse onto only $\sim\!42$ distinct frequency peaks because each single-particle move contributes the same $\omega=\varepsilon_a-\varepsilon_b$ for all $2^{N-2}$ spectator configurations (SM~\cite{suppmat}). The GUE reservoir, by contrast, has Haar-random eigenvectors that lack this Slater-determinant structure, so all $65{,}280$ transitions acquire nonzero weight and form a featureless quasi-continuum [Fig.~\ref{fig:fig1}(d)].

To characterize this frequency-selective filtering we use two descriptive structural proxies computed from the undriven $H_0$ eigenstates, the transition kernel $A_H(\omega)$ [Eq.~\eqref{eq:AH}] and its overlap $\Omega$ with the class-discriminative signal spectrum (definitions and per-phase values in SM~\cite{suppmat}). These proxies are descriptive rather than predictive. Although the GUE reservoir has the largest total coupling weight and $\Omega_\text{GUE}>\Omega_\text{crit}$, it performs significantly worse, and the localized phase has the highest per-feature Fisher discriminability yet not the highest accuracy. It is thus the \emph{distribution} of spectral weight (concentrated peaks vs.\ a quasi-continuum), rather than the total overlap, that tracks performance (SM~\cite{suppmat}).

\begin{figure}[tb]
  \centering
  \includegraphics[width=\columnwidth]{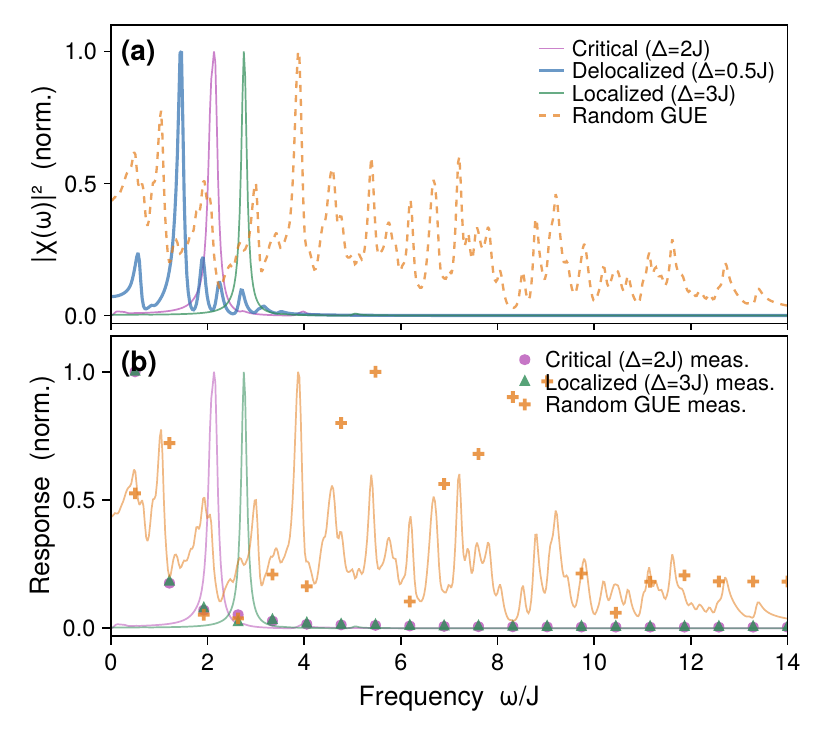}
  \caption{(a)~Hamiltonian transition kernel $|\chi(\omega)|^2$ (normalized) for all
  four reservoir types, computed from undriven $H_0$ eigenstates and used here as a
  structural descriptor of the closed-system response (not as the full Liouvillian
  susceptibility).  AAH reservoirs show sharp peaks at low frequencies,
  whereas the GUE response is broad and featureless.
  (b)~Structural kernel $|\chi(\omega)|^2$ (lines) and the empirically measured weak-probe
  transfer function $G(\omega_d)$ (symbols; $\kappa|s|\approx 0.02J$). The measured AAH response
  is concentrated at low frequencies ($\omega/J\lesssim 2$) while the GUE response is broad and
  extends to high frequencies. We use $|\chi(\omega)|^2$ as a structural descriptor of the
  undriven basis, not as a quantitative model of the driven transfer function.}
  \label{fig:kubo}
\end{figure}

We validate this picture with two complementary quantities, the Lorentzian-broadened transition kernel $|\chi(\omega)|^2$ computed from the undriven $H_0$ eigenstates and the windowed power spectral density of the driven observable,
\begin{equation}\label{eq:psd}
  P(\omega) = \frac{1}{T_w}\left| \int_{t_0}^{t_0+T_w} \langle O(t) \rangle e^{-i\omega t} dt \right|^2\,,
\end{equation}
with $O=\hat{n}_1$ and $T_w=500$ steps after a 150-step washout. Figure~\ref{fig:kubo}(a) shows that the AAH reservoirs exhibit sharp resonance peaks at low frequencies ($\omega/J \lesssim 4$), concentrated near the discriminative SNA frequencies $\omega_1,\omega_2 \sim 1$--$2\,J$, whereas the GUE response is broad and featureless out to $\omega/J \sim 12$. A weak-probe measurement of the transfer function $G(\omega_d)$ (effective drive $\kappa|s|\approx 0.02J$) confirms this qualitative separation, with the AAH response concentrated at low frequencies ($\omega/J\lesssim 2$) and the GUE response spread across a broad high-frequency band [Fig.~\ref{fig:kubo}(b)]; we use $|\chi(\omega)|^2$ as a structural descriptor of the undriven basis rather than a quantitative predictor of the driven response; the reservoir's behavior under strong drive ($\kappa|s|\sim 2J$) is characterized separately through the nonlinear PSD (state-dressing analysis in SM~\cite{suppmat}, Fig.~\ref{fig:S1_nonlinear}).

To separate spectral sparsity from golden-ratio matching and from AAH's specific eigenvectors, we run controlled experiments at $\Delta=2J$, $N=8$ (SM~\cite{suppmat}, Tables~\ref{tab:ablation} and~\ref{tab:causal}). Replacing $\beta=\varphi$ by a mismatched irrational or a rational approximant leaves the peak accuracy near $90\%$, well above GUE ($83.5\%$, $p<10^{-4}$), so golden-ratio matching is not the primary determinant and confers only a short-$T$ convergence-speed advantage ($89.8\%$ vs.\ $84.5\%$ at $T=50$). The mismatched case in fact yields a higher raw $\Omega$ without improved accuracy.

Two symmetry interventions isolate the mechanism. Random on-site disorder retains the free-fermion quadratic structure and the sparse 3584-transition pattern while randomizing frequencies, and it matches AAH ($89.9\%$ vs.\ $89.8\%$, the deterministic AAH value lying within the disorder ensemble confidence interval). Permuting the eigenvectors instead preserves the AAH spectrum but destroys the Slater-determinant selection rules, activating all $65{,}280$ transitions and collapsing accuracy to the dense-reservoir level ($79.7\%$, GUE $82.3\%$). Free-fermion sparsity thus accounts for the advantage over dense reservoirs. The permutation also delocalizes the coupling operator, a residual confound noted in the SM.

The mechanism is not specific to the Duffing oscillator. On the structurally distinct quasiperiodically forced logistic map (200 Lyapunov-verified signals; SM~\cite{suppmat}), the key findings reproduce: all AAH variants outperform GUE ($p<10^{-4}$), the rational approximant is indistinguishable from the control ($p=0.55$), and the $\beta$-mismatch falls slightly below the control ($95.0\%$ vs.\ $97.3\%$, $p=0.013$) while remaining well above GUE ($90.1\%$).

The controlled experiments above, on systems with known labels, establish that the reservoir separates SNA from chaos. As a training-free application requiring no labeled data, we then use it to screen real observational time series: the Kepler RR Lyrae variable stars of Lindner~\textit{et al.}~\cite{Lindner2015}, whose ``golden'' RRc members have pulsation-frequency ratio $f_2/f_1\approx\varphi^{-1}$ and were argued to host strange-nonchaotic dynamics. Driving the reservoir with each light curve and ranking the resulting feature vector against phase-randomized surrogates, the screen flags one golden star, KIC~4064484, as a candidate departing the surrogate null and warranting dedicated dynamical study. The full surrogate analysis for all seven targets is given in the SM~\cite{suppmat}, using a reservoir of only $256$ Hilbert-space dimensions ($N=8$).

For spectrally structured tasks, our results indicate that performance depends on which Fock-space transitions carry weight rather than on proximity to a quantum phase transition alone. The sparse free-fermion transition set acts as a frequency-selective filter for the singular-continuous spectra that separate SNA from chaos. Because a quadratic free-fermion reservoir is efficiently simulable classically, this result is not a quantum speedup; it is a spectral-matching principle that favors sparse Hamiltonians compatible with near-term hardware~\cite{Roushan2017} and parallels frequency-selective classical delay-line reservoirs~\cite{Appeltant2011,Larger2012}. The training-free readout also permits surrogate-based screening of experimental time series, as illustrated with the Kepler light curves.

One caveat bounds the mechanism. The two accurate reservoirs, AAH and random on-site disorder, are both quadratic and $U(1)$ particle-conserving, whereas the two inaccurate reservoirs break both properties simultaneously. Our controls therefore establish free-fermion one-body transition sparsity as the \emph{dominant observed factor} without isolating it from the conservation law that generates it. Full separation remains open and will require particle-conserving interacting or within-sector controls, together with aperiodic (Fibonacci, Thue-Morse) reservoirs that decouple sparsity from quasiperiodic order, as well as a predictive replacement for the descriptive transition-kernel overlap $\Omega$~\cite{suppmat}. The amplitude-control analysis (SM~\cite{suppmat}, Table~\ref{tab:amp}) and the second-system benchmark (Fig.~\ref{fig:qplogistic}) retain the sparse-over-dense advantage, arguing against an amplitude artifact or a dataset-specific cue.

The controlled benchmarks support transition-spectrum engineering as a reservoir-design strategy. A Hamiltonian with a sparse transition kernel aligned with the target dynamics provides a compact processor for signals that are difficult to distinguish using conventional dynamical invariants. Implementing such reservoirs on programmable quantum hardware would test this strategy experimentally.



\clearpage
\onecolumngrid

\setcounter{equation}{0}
\setcounter{figure}{0}
\setcounter{table}{0}
\renewcommand{\theequation}{S\arabic{equation}}
\renewcommand{\thefigure}{S\arabic{figure}}
\renewcommand{\thetable}{S\arabic{table}}

\begin{center}
  {\large\textbf{Supplemental Material}}\\[4pt]
  \textit{Free-fermion spectral structure enables strange nonchaotic attractor classification
  in Aubry-Andr\'{e}-Harper quantum reservoirs}
\end{center}

\vspace{6pt}

The power spectrum of a strange nonchaotic attractor is neither a small set of
tones, as in periodic or quasiperiodic motion, nor a smooth broadband signal, as
in chaos or noise. It is a self-similar comb of peaks across many scales.
Lyapunov exponents do not separate these nonchaotic signals from regular motion,
although their spectra can be as complex as chaotic spectra. The quantum
reservoir acts as a bank of frequency filters: its response is strongest near
its internal energy-level transitions, and its output samples the signal power
at those frequencies. In the Aubry-Andr\'{e}-Harper reservoir, the active
transitions form a sparse set rather than a dense continuum. The quadratic
free-fermion structure leaves only a small fraction of transitions active, while
the critical spectrum has Cantor-set structure. This produces a probe comb that
selectively responds to the spectral feature separating strange-nonchaotic
signals from chaos. The controlled experiments below attribute the classification
advantage to the sparsity of the active transitions, not to proximity to the
phase transition itself.

\section{Strange nonchaotic attractors: physics and discrimination challenge}

Strange nonchaotic attractors (SNAs) were first described by Grebogi,
Ott, Pelikan, and Yorke~\cite{Grebogi1984} and arise generically in
quasiperiodically forced dissipative systems.  Their defining characteristic
is a combination that appears paradoxical: the attractor has fractal
(non-integer Hausdorff dimension, non-differentiable) geometry, yet all
Lyapunov exponents are nonpositive.  In conventional dynamical systems
theory, fractal geometry is associated with chaos, but SNAs demonstrate
that fractal structure is possible without exponential sensitivity to initial
conditions.

\subsection{The quasiperiodically forced Duffing oscillator}

We generate discrimination signals from the quasiperiodically forced Duffing
oscillator:
\begin{equation}
  \ddot{x} + \delta\dot{x} + \alpha x + \beta_D x^3
  = A_1\cos(\omega_1 t) + A_2\cos(\omega_2 t)\,,
\label{eq:duffing}
\end{equation}
with parameters $\delta=0.2$, $\alpha=-1$, $\beta_D=1$, $\omega_1=1$,
$A_1=0.3$, and incommensurability ratio $\omega_2/\omega_1=\varphi=(\sqrt{5}-1)/2$.
The forcing frequencies are in the golden ratio, ensuring that the system
is quasiperiodically driven. The driving is not periodic, but its spectral
content lies at integer linear combinations $n\omega_1 + m\omega_2$
with $n,m\in\mathbb{Z}$.

\begin{figure}[htb]
  \centering
  \includegraphics[width=0.85\textwidth]{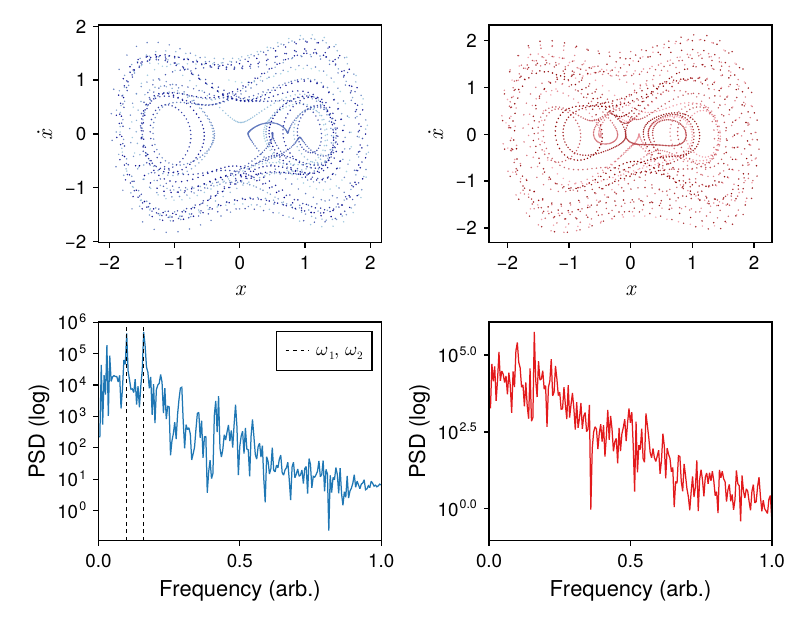}
  \caption{Quasiperiodically forced Duffing oscillator.
  (a,b)~Phase portraits $(x, \dot{x})$ for SNA ($A_2=0.50$,
  $\lambda_\text{max}<0$) and chaos ($A_2=0.78$, $\lambda_\text{max}>0$).
  Despite their qualitatively similar appearance, the two regimes are
  distinguished by their spectral content [panels (c,d)]: the SNA retains
  pronounced quasiperiodic peaks at $n\omega_1+m\omega_2$ (dashed lines mark
  $\omega_1$, $\omega_2$) on a singular-continuous background, while chaos
  produces a broadband absolutely continuous spectrum. Panels (a,b) and
  (c,d) of the main text Fig.~1 reproduce the spectra side-by-side with
  the reservoir transition spectra $A_H(\omega)$ to highlight the
  sparse-versus-dense parallel.}
  \label{fig:sm_sna_phase}
\end{figure}

By varying the forcing amplitude $A_2$, the system traverses a
transition from SNA to chaos:
\begin{itemize}
  \item \textbf{SNA regime} ($A_2\in[0.42,0.60]$): The maximal Lyapunov
  exponent $\lambda_\text{max}<0$.  The attractor is geometrically strange
  (fractal in the phase-frequency plane) but dynamically stable.  The power
  spectrum is typically singular-continuous, with pronounced peaks at
  quasiperiodic frequencies $n\omega_1+m\omega_2$ (indexed by the ring
  $\mathbb{Z}[\varphi]$) embedded in a fractal background.
  \item \textbf{Chaotic regime} ($A_2\in[0.68,0.90]$): $\lambda_\text{max}>0$.
  Sensitivity to initial conditions broadens the spectral peaks into a
  continuous background.  The quasiperiodic fine structure is destroyed.
\end{itemize}
We sample 100 signals per class (total 200), each of length 1500 time steps
with effective step $\Delta t_\text{eff} = 5\times 0.05 = 0.25$, after a
transient of 500 steps.  Representative phase portraits and the corresponding
power spectra are shown in Fig.~\ref{fig:sm_sna_phase}; we note that the two
classes occupy disjoint forcing-amplitude intervals, so part of the spectral
distinction reflects an amplitude-dependent bandwidth difference rather than
a pure SNA--chaos topological distinction.

\subsection{Why discrimination is hard}

The difficulty of SNA detection stems from the following.  Standard
chaos indicators such as correlation dimension $D_2$, Lyapunov spectrum, and recurrence
plots all produce ambiguous results near the SNA--chaos boundary because:
(i)~both attractors are geometrically complex; (ii)~the Lyapunov exponent
of the SNA is zero or weakly negative, barely distinguishable from chaos in
finite data; (iii)~both classes are non-periodic.  The most reliable
discriminant is the spectral fine structure: SNAs have power concentrated
at $\mathbb{Z}[\varphi]$ frequencies while chaos has a continuous spectrum.
But detecting this distinction requires a feature extractor sensitive to
the \emph{arithmetic organization} of the spectral content, not just its
gross power.

\section{Quantum reservoir computing: architecture and training}

\subsection{Lindblad dynamics and virtual nodes}

The quantum reservoir is governed by the Lindblad master equation
\begin{equation}
  \dot{\rho} = \mathcal{L}[\rho]
  \equiv -i[H_0 + H_\text{in}(t),\,\rho]
  + \sum_k \gamma_k\!\left(L_k\rho L_k^\dagger
    - \tfrac{1}{2}\{L_k^\dagger L_k,\,\rho\}\right),
\end{equation}
where the Lindblad jump operators implement three decoherence channels:
\begin{align}
  &\text{Dephasing:}\quad L_j^\text{deph} = \sqrt{\gamma_\text{deph}}\,\hat{n}_j,
  \quad \gamma_\text{deph} = 0.05,\\
  &\text{Damping:}\quad L_j^\text{damp} = \sqrt{\gamma_\text{damp}}\,c_j,
  \quad \gamma_\text{damp} = 0.02,\\
  &\text{Pumping:}\quad L_j^\text{pump} = \sqrt{\gamma_\text{damp}\bar{n}}\,c_j^\dagger,
  \quad \bar{n} = 0.1.
\end{align}
These rates are chosen to be weak compared to the hopping $J$ ($\gamma/J \ll 1$), so the
dissipation broadens the effective linewidths without washing out the
Hamiltonian structure.  In physical units, with $J\sim 100$\,MHz, these correspond to $T_2 = 1/\gamma_\text{deph} = 200$\,ns and $T_1 = 1/\gamma_\text{damp} = 500$\,ns, compatible with the coherence-to-hopping ratio of early-generation superconducting qubit simulators~\cite{Roushan2017}.  Modern transmon devices achieve substantially longer absolute $T_1, T_2$ (up to the millisecond range), corresponding to an even more favorable dissipation regime in our framework.  Varying $\gamma_\text{deph}\in[0.025, 0.075]$ and $\gamma_\text{damp}\in[0.01, 0.03]$ changes the best accuracy by less than $3$ percentage points for all AAH phases (Fig.~\ref{fig:robustness}).  The master equation is integrated numerically with step $\delta t = 0.05$.

The time-multiplexing scheme creates $M=5$ virtual nodes per input step:
each input value $s_k$ is held constant while the system evolves for
$M\delta t = 0.25$ time units.  This effectively increases the feature
dimensionality by a factor of $M$ without changing the physical system,
a standard technique in reservoir computing~\cite{Nakajima2019}.

\subsection{Feature extraction and readout}

After a washout period of 200 steps (discarded to eliminate transient effects),
we record the following observables at each virtual node:
\begin{itemize}
  \item Site occupancies: $\langle\hat{n}_j\rangle = \mathrm{tr}(\hat{n}_j\rho)$
  for $j=1,\ldots,8$ (8 observables per node).
  \item Nearest-neighbor correlators:
  $\langle\hat{n}_j\hat{n}_{j+1}\rangle$ for $j=1,\ldots,7$ (7 observables per node).
\end{itemize}
This gives $15\times 5 = 75$ raw observables per time step.  Over the full
post-washout window of $T$ steps, we compute the mean and standard deviation
of each observable, yielding $2\times 75 = 150$ final observables per signal.
Mean pooling captures the time-averaged spectral content, whereas standard deviation
pooling captures the fluctuation amplitude.

The readout is a regularized linear regression:
\begin{equation}
  \hat{y} = \mathbf{x}^\top \mathbf{W}_\text{out},\qquad
  \mathbf{W}_\text{out} = (X^\top X + \lambda I)^{-1} X^\top \mathbf{y},
\end{equation}
with regularization parameter $\lambda=10^{-4}$.  Classification is performed by
thresholding $\hat{y}$ at a fixed threshold of $0.5$ (SNA: $y=0$, chaos: $y=1$); the threshold is not optimized on the training set.  Accuracy is
averaged over 10 random 70/30 stratified train--test splits.

The full accuracy curves versus processing horizon $T$ are shown in
Fig.~2 of the main text.

\section{Lyapunov exponent regression}

\begin{figure}[htb]
  \centering
  \includegraphics[width=0.5\textwidth]{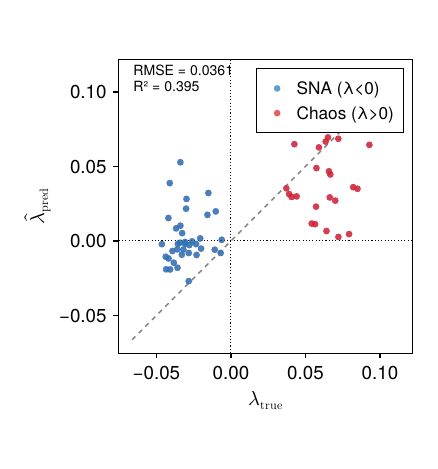}
  \caption{Predicted vs true $\lambda_\text{max}$ for the critical-phase
  QRC reservoir.  Dashed diagonal: perfect regression.  Blue: SNA signals
  ($\lambda<0$); red: chaotic signals ($\lambda>0$).  The QRC correctly
  separates the two classes and captures the ordering of Lyapunov exponents
  within each class ($R^2=0.395$), demonstrating that the reservoir encodes
  quantitative dynamical information beyond a binary label.}
  \label{fig:lyap_reg}
\end{figure}

Beyond binary discrimination, we test whether the QRC can regress the
\emph{value} of the maximal Lyapunov exponent $\lambda_\text{max}$.  This is a
harder task: the readout must encode a continuous quantity, not a binary label,
and must generalize across the SNA--chaos boundary where $\lambda_\text{max}
\approx 0$.  Using the same feature extraction methodology and ridge regression
with the critical-phase reservoir ($\Delta=2J$, $N=8$), we obtain $R^2=0.395$
and RMSE~$\approx 0.036$ (Fig.~\ref{fig:lyap_reg}).

While the $R^2$ is modest, the result is significant for two reasons.  First,
the reservoir recovers the \emph{sign} of $\lambda_\text{max}$ (i.e., classifies
SNA vs.\ chaos) with high confidence, as expected from the binary discrimination
results.  Second, the approximate ordering of $\lambda_\text{max}$ values
within the chaotic class (larger $A_2$ gives more positive $\lambda$) is also
recovered, indicating that the reservoir is sensitive to the \emph{degree} of
chaos, not just its presence.  This is consistent with the spectral resonance
picture: more strongly chaotic signals have more broadband spectra and are
therefore more easily distinguished from SNAs by the spectrally matched reservoir.

\section{Aubry-Andr\'{e}-Harper spectrum: level statistics and IDOS}

\subsection{Energy level statistics and localization transition}

The AAH model defined in Eq.~\eqref{eq:AAH} exhibits a sharp metal-insulator transition at the critical value $\Delta_c=2J$. This transition is reflected in the single-particle level spacing distribution $P(s)$, which changes character across the phases of the model.

Figure~\ref{fig:level_spacing}(a) shows $P(s)$ obtained by pooling
single-particle spacings from $N=200$-site AAH chains over $50$ random
initial-phase realizations $\phi_0$ (bulk levels only, edge levels removed
to suppress band-edge distortions). For the delocalized phase ($\Delta=0.5J$), $P(s)$ exhibits finite-size level repulsion (small-$s$ suppression) and a distribution peak near $s\sim 0.9$, broadly consistent in shape with the Wigner surmise $P(s)\approx \frac{\pi}{2}s\,e^{-\pi s^2/4}$. We emphasize that this resemblance is descriptive only: the AAH spectrum is deterministic and quasiperiodic and is not GOE/GUE-universal, so it does not satisfy random-matrix theory strictly. We use ``finite-size level repulsion'' rather than ``Wigner-Dyson statistics'' as the precise characterization. The localized phase ($\Delta=3J$) shows a markedly different distribution, with weight shifted toward small $s$ and an exponential tail consistent with the approach to Poisson statistics in the localized regime. The critical phase ($\Delta=2J$) exhibits a sharp peak at $s\!\to\!0$, the signature of the Cantor-set/multifractal spectrum: many levels are tightly clustered within the self-similar gap hierarchy, producing a large fraction of small spacings that neither level-repulsion nor Poisson statistics describes.

Despite the qualitative GOE-like statistics in the delocalized phase, the delocalized AAH phase is qualitatively distinct from GUE in its \emph{eigenvector} structure: AAH eigenstates are Bloch-like extended states with spatially structured amplitudes, whereas GUE eigenvectors are uniformly random. This structured eigenvector organization creates the free-fermion selection rules that render ${\sim}94.5\%$ of Fock-space transitions inactive under the local coupling $\hat{n}_1$, explaining why the delocalized AAH phase substantially outperforms GUE despite both having extended eigenstates. It is well-established that the many-body Fock-space spectrum of non-interacting systems reflects these single-particle Anderson localization properties~\cite{Anderson1958}, and the level-statistics crossover is a standard diagnostic for the localization transition~\cite{Evers2008}.

\subsection{Integrated density of states and the gap labeling theorem}

A fundamental property of the AAH model is described by its integrated density of states (IDOS), $\mathcal{N}(E)$. According to the gap labeling theorem~\cite{Bellissard1982}, for an irrational incommensurability parameter $\beta$, the values of $\mathcal{N}(E)$ within the spectral gaps are restricted to the module $\mathbb{Z}+\mathbb{Z}\beta$. In the present case with $\beta=\varphi$, this module corresponds to the ring of integers $\mathbb{Z}[\varphi]$. The resulting IDOS takes the form of a devil's staircase, which increases only on the support of the Cantor-set spectrum and remains constant within the gaps.

\begin{figure}[htb]
  \centering
  \includegraphics[width=0.85\textwidth]{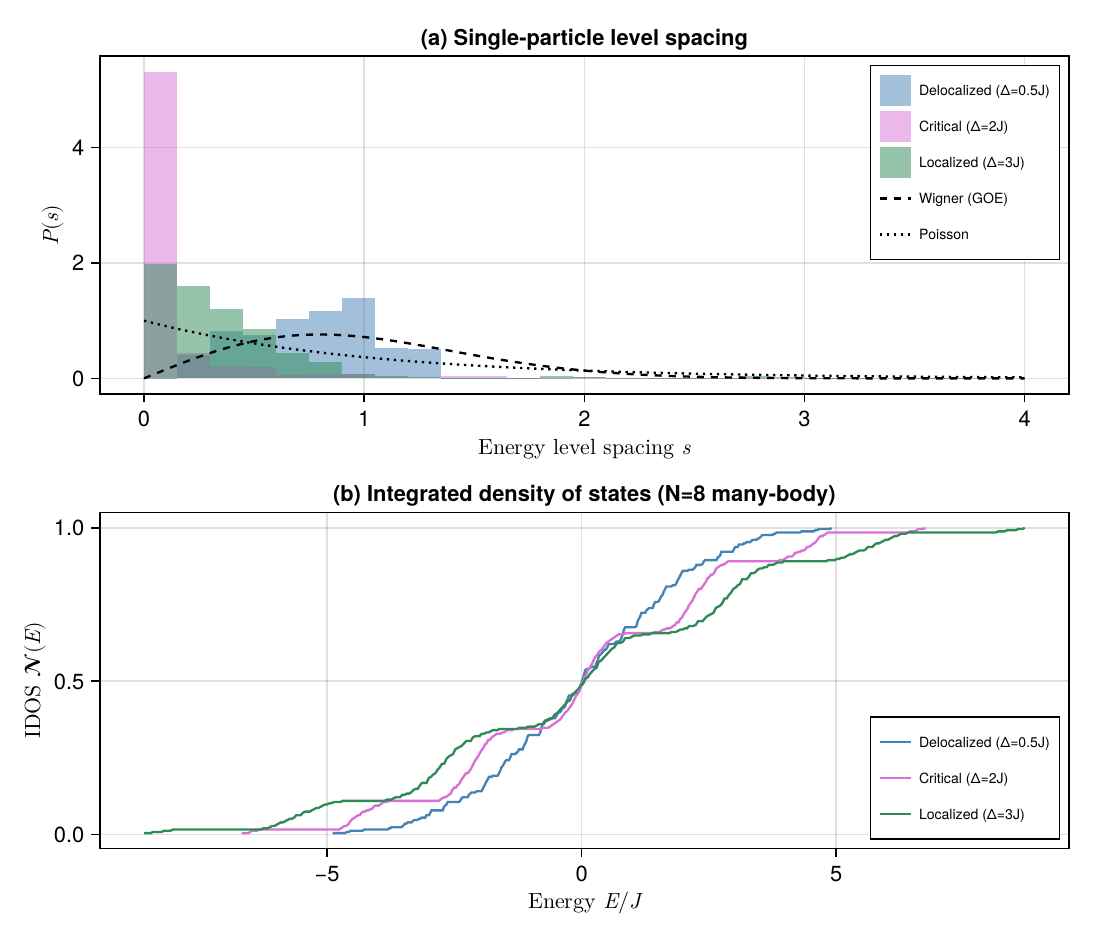}
  \caption{(a)~Single-particle nearest-neighbor level spacing distribution
  $P(s)$ for the three AAH phases, pooled over $50$ random initial-phase
  realizations of $N=200$-site chains (bulk levels only).  Delocalized
  (blue, $\Delta=0.5J$) shows finite-size level repulsion broadly consistent
  in shape with the Wigner surmise (dashed) but not random-matrix-universal;
  localized (green, $\Delta=3J$) shows
  weight shifted toward small $s$, approaching Poisson statistics (dotted);
  critical (pink, $\Delta=2J$) exhibits a sharp peak at $s\!\to\!0$
  reflecting the Cantor-set / multifractal spectrum at the metal-insulator
  transition.
  (b)~Many-body integrated density of states $\mathcal{N}(E)$ for $N=8$
  qubits, showing devil's-staircase structure.  The critical phase has the
  richest staircase, with plateaus at gap values $\in\mathbb{Z}[\varphi]$
  guaranteed by the gap labeling theorem~\cite{Bellissard1982} and the
  Avila-Jitomirskaya (Ten Martini) theorem~\cite{Avila2009}.}
  \label{fig:level_spacing}
\end{figure}

At the critical point ($\Delta=2J$), the Avila-Jitomirskaya theorem~\cite{Avila2009} ensures that all gaps predicted by the labeling theorem are indeed open, resulting in a fully developed self-similar Cantor set. In the other phases, some of these gaps may close, which results in a simpler staircase structure. The IDOS curves for the three phases are presented in Fig.~\ref{fig:level_spacing}(b). The staircase feature is most prominent in the critical phase, whereas the other phases exhibit smoother curves due to gap closures.

The connection to machine-learning performance is suggestive rather than
deductive. Strictly, IDOS gap labels $p+q\beta$ are dimensionless quantum
numbers that label \emph{plateaux} of $\mathcal{N}(E)$, not values of the
single-particle gaps $E_{i+1}-E_i$ themselves, and they are still further
removed from the many-body transition frequencies $\omega_{mn}=E_m-E_n$
that enter $A_H(\omega)$. We use the coincidence of the modules
$\mathbb{Z}+\mathbb{Z}\beta$ (gap labels) and $\mathbb{Z}+\mathbb{Z}\varphi$
(SNA forcing) at $\beta=\varphi$ as a heuristic indicator of arithmetic
compatibility between reservoir and signal, supplemented by the direct
numerical peak-overlap analysis in the main text [Fig.~\ref{fig:fig1},
Fig.~\ref{fig:kubo}].

\section{Memory capacity and Fisher discriminability}

\subsection{Linear memory capacity}

The memory capacity of a reservoir is defined by the following relation:
\begin{equation}
  C_k = \frac{[\mathrm{cov}(\hat{s}_{t-k}, s_{t-k})]^2}
             {\mathrm{var}(\hat{s}_{t-k})\,\mathrm{var}(s_{t-k})}\,,
\end{equation}
where $\hat{s}_{t-k}$ denotes the optimal linear reconstruction of the input $s$ at a delay of $k$ steps, based on the current reservoir state. The total memory capacity is obtained by summing over all delays, $C = \sum_{k=1}^{\infty} C_k$. For a linear system with $D$ degrees of freedom, the theoretical limit is $C \leq D$; in the present work, with 150 observables, the maximum possible capacity is 150.

We evaluate $C_k$ for delays $k=0,1,\ldots,30$ using a standard white-noise input protocol (i.i.d.\ uniform $s_t\in[-1,1]$), following Jaeger~\cite{Jaeger2002}. Results are shown in Fig.~\ref{fig:memory_fisher}(a), including the instantaneous response $C_0$. The total capacity $C=\sum_{k=0}^{30}C_k$ lies in the range $1.1$--$1.5$ for all investigated phases, far below the 150-observable theoretical maximum and the $C\approx 50$ typical of classical echo-state networks. This confirms that the standard linear memory metric is unreliable in the strongly nonlinear driving regime ($\kappa=2.0J$); the reservoir operates in a regime where the linear-response assumptions underlying the memory capacity definition break down, rather than being literally memoryless.

This observation is physically consistent with the dissipation rates used in the Lindblad equation: the dephasing and damping time scales ($\sim 20$ and $50$ units, respectively) are short compared to the processing horizon $T$. The observed $T$-dependence of accuracy (Fig.~2 of the main text) arises not from temporal memory but from the statistics of mean-std pooling: at small $T$, the pooled features have high sampling variance; at large $T$, discriminative features are diluted by averaging over long windows. We verify this by computing the per-feature Fisher ratio $\bar{F}(T)$ as a function of $T$ (Fig.~\ref{fig:fisher_vs_T}, below). The delocalized phase shows the expected non-monotonic profile with a peak near $T\!=\!100$, mirroring the accuracy curve in Fig.~2 of the main text; the critical phase declines monotonically (its accuracy peak at $T\!=\!50$ lies at the start of the plotted range); and the localized phase varies more weakly with $T$, consistent with the broader feature plateau in Fig.~2 across $T\!=\!200$--$700$.

\subsection{Fisher discriminability}

\begin{figure}[htb]
  \centering
  \includegraphics[width=0.85\textwidth]{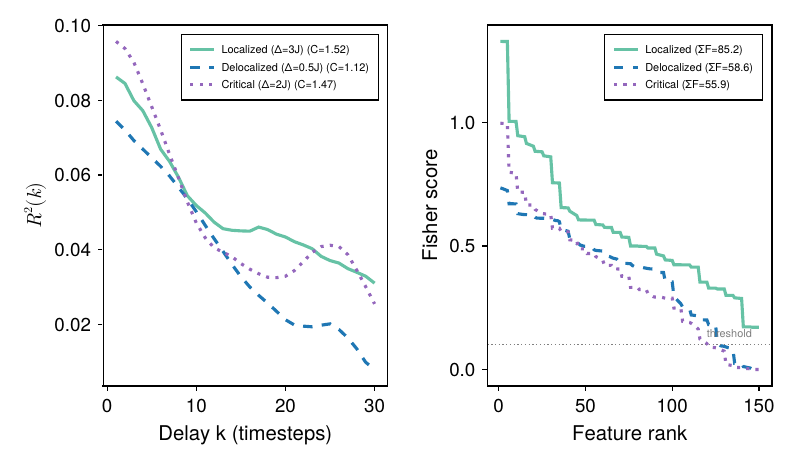}
  \caption{(a)~Memory capacity $R^2(k)$ vs delay $k$ for the three AAH phases, measured with white-noise input.
  Total capacity $C\approx 1.1$--$1.5$ reflects the breakdown of linear memory metrics in the strongly nonlinear driving regime.
  (b)~Sorted Fisher ratios for all 150 observables at $T=T_\text{peak}$.  The localized phase (green)
  has the highest total discriminability ($\Sigma F=85.2$, all 150 observables
  above the significance threshold $F=0.1$, dotted line); delocalized (blue) and critical (purple) have $\Sigma F=58.6$ and $55.9$ respectively.
  The Fisher-ratio dependence on $T$ is shown separately in Fig.~\ref{fig:fisher_vs_T}.}
  \label{fig:memory_fisher}
\end{figure}

\begin{figure}[htb]
  \centering
  \includegraphics[width=0.6\textwidth]{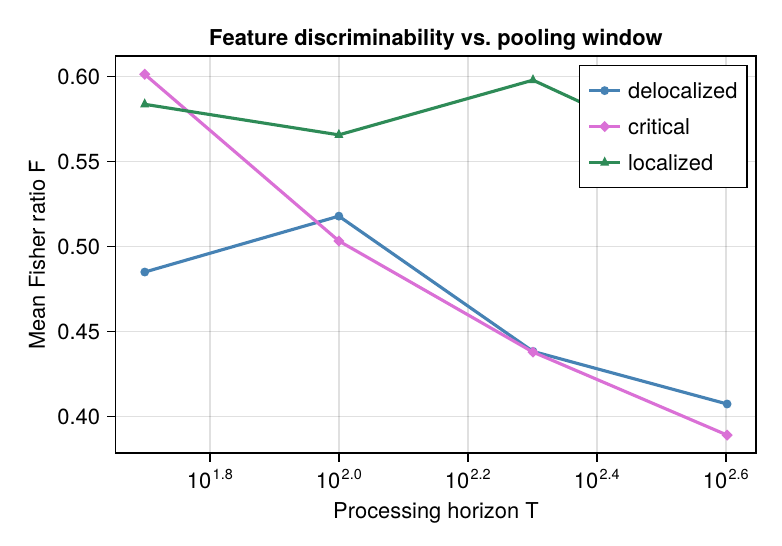}
  \caption{Mean Fisher ratio $\bar{F}$ as a function of the processing horizon $T$ for the three AAH phases.  Delocalized (blue) peaks near $T=100$; critical (pink) declines monotonically from the peak at $T=50$; localized (green) is approximately flat over the plotted range.  These trends are consistent with the phase-dependent best-$T$ values in Fig.~2 of the main text.}
  \label{fig:fisher_vs_T}
\end{figure}

The discriminative power of individual observables is assessed using the Fisher ratio, defined as:
\begin{equation}
  F_i = \frac{(\mu_i^\text{SNA} - \mu_i^\text{chaos})^2}
             {(\sigma_i^\text{SNA})^2 + (\sigma_i^\text{chaos})^2}\,,
\end{equation}
where $\mu_i$ and $\sigma_i$ represent the class-specific mean and standard deviation for the $i$-th feature. A high value of $F_i$ indicates that the feature effectively separates the SNA and chaotic classes. The sorted Fisher ratios for all 150 observables are presented in Fig.~\ref{fig:memory_fisher}.

The localized phase exhibits the highest total discriminability ($\Sigma F = 85.2$, with all observables exceeding the significance threshold of 0.1). This is attributed to its sharp and isolated resonance peaks, which respond selectively to the quasiperiodic frequencies of the SNA signal. In contrast, the delocalized and critical phases show lower total discriminability ($\Sigma F = 58.6$ and $55.9$, respectively), as their broader spectral responses lead to a reduction in per-feature selectivity.

The localized phase thus has the highest $\Sigma F$ yet not the highest classification accuracy ($93.7\%$ vs.\ $94.7\%$ for the delocalized phase), because the Fisher ratio is a \emph{univariate} metric that assesses each feature independently, whereas the ridge regression readout exploits multivariate correlations. The delocalized phase produces features with stronger inter-feature correlations (effective rank of the feature covariance matrix: $r_\text{eff}=1.8$ vs.\ $1.6$ for the localized phase), enabling the linear readout to construct more effective discriminant functions despite lower per-feature separability.

\section{Amplitude control}

Because the SNA and chaotic classes are generated at disjoint forcing amplitudes
($A_2\in[0.42,0.60]$ vs.\ $[0.68,0.90]$), we quantify the class information
carried by signal amplitude alone and compare it with the reservoir's performance
on amplitude-matched inputs. All inputs are min-max normalized to $[-1,1]$; the
per-signal standard deviation nevertheless remains systematically larger for SNA
than for chaos signals ($0.521\pm0.007$ vs.\ $0.497\pm0.009$), and a one-parameter
threshold classifier on this single feature, with threshold and sign fitted on
each training split, reaches $89.7\pm2.5\%$ (Table~\ref{tab:amp}).

We then rescale every signal to zero mean and exactly unit variance, removing
per-signal amplitude and energy as cues, and repeat the classification with the
identical reservoir ($\Delta=2J$, $N=8$), ridge
readout, and protocol (all $n=100$ signals per class, 20 stratified $70/30$ splits,
best accuracy over $T\in\{50,100,200,400\}$). On the manuscript signals this
protocol reproduces the main result ($89.7\pm3.0\%$ at $T=50$; cf.\
$89.8\pm2.5\%$ in Table~\ref{tab:causal}). On the variance-matched signals the
accuracy is fully retained, in fact slightly higher ($93.2\pm2.6\%$ at
$T=50$), even though the amplitude cue is absent by construction. The reservoir's
discrimination therefore rests on the signals' spectral structure, not on their
forcing amplitude.

\begin{table}[htb]
  \centering
  \caption{Amplitude control for the Duffing task ($\Delta=2J$, $N=8$, $n=100$ per
  class, 20 stratified $70/30$ splits, best accuracy over $T\in\{50,100,200,400\}$).
  Rescaling every signal to unit variance removes per-signal amplitude as a cue;
  the reservoir accuracy is fully retained.}
  \label{tab:amp}
  \begin{tabular}{@{}lc@{}}
    \hline\hline
    Classifier / input & Accuracy (\%) \\
    \hline
    Amplitude-only (per-signal std, fitted threshold) & $89.7\pm2.5$ \\
    Reservoir, manuscript signals & $89.7\pm3.0$ \\
    Reservoir, unit-variance-matched & $93.2\pm2.6$ \\
    \hline\hline
  \end{tabular}
\end{table}

To confirm that the sparse-reservoir \emph{advantage}, and not merely the AAH
accuracy, survives removal of the amplitude cue, we repeat the four-way ablation
(AAH, on-site disorder, GUE, eigenvector-permutation) on the variance-matched
signals (Table~\ref{tab:amp_ablation}). Here all four Hamiltonians are evaluated
under an identical reservoir template (the AAH dissipation and coupling) and a
reduced processing-horizon sweep $T\in\{50,100\}$ on a representative
$50$-per-class stratified subsample with four seeds; absolute margins therefore
differ from the main-text intervention, but the comparison is internally
consistent and isolates the effect of variance matching. On the original signals
the two sparse free-fermion reservoirs lead the two dense reservoirs by a modest
margin. Removing the amplitude cue does not close this gap; it \emph{widens} it.
The AAH advantage over GUE grows from $1.9$ to $6.5$ percentage points, with
nearly non-overlapping seed ranges (AAH $90.5$--$91.8\%$ vs GUE
$82.3$--$87.0\%$), and the eigenvector-permutation control remains at the
dense-reservoir baseline ($85.8\%$, versus AAH $91.2\%$). The reservoir advantage
therefore reflects spectral structure rather than the amplitude cue, which if
anything masks part of the separation on the raw signals.

\begin{table}[htb]
  \centering
  \caption{Four-way ablation on original versus unit-variance-matched signals
  (representative $50$-per-class stratified subsample, $4$ seeds, best accuracy
  over $T\in\{50,100\}$; mean$\pm$s.d.\ over seeds, all reservoirs under a common
  dissipation template). The AAH-over-GUE margin \emph{widens} under variance
  matching and the eigenvector-permutation control stays at the dense baseline,
  confirming that the sparse-reservoir advantage is not an amplitude artifact.}
  \label{tab:amp_ablation}
  \begin{tabular}{@{}lcc@{}}
    \hline\hline
    Reservoir & Original (\%) & Variance-matched (\%) \\
    \hline
    AAH (sparse)                 & $87.1\pm1.1$ & $91.2\pm0.5$ \\
    On-site disorder (sparse)    & $89.5\pm3.6$ & $88.3\pm2.0$ \\
    GUE (dense)                  & $85.2\pm1.0$ & $84.7\pm1.9$ \\
    Eigenvector-permuted (dense) & $85.3\pm1.4$ & $85.8\pm1.0$ \\
    \hline\hline
  \end{tabular}
\end{table}

\section{Spectral overlap theory: derivation and numerical results}

\subsection{Transition-frequency spectrum}

The response of the quantum reservoir to an input perturbation is governed by
its transition-frequency spectrum.  We define the weighted spectral function
\begin{equation}
  A_H(\omega) = \sum_{m\neq n}|\langle m|V|n\rangle|^2\,\delta(\omega-\omega_{mn}),
\end{equation}
where $V=\kappa\hat{n}_1$ is the input coupling operator,
$\omega_{mn}=(E_m-E_n)/\hbar$, and $|m\rangle$ are eigenstates of $H_0$.
For numerical evaluation, the delta functions are replaced by Lorentzians with
broadening $\gamma=0.07J$ (matching the total dissipative linewidth,
$\gamma_\text{deph}+\gamma_\text{damp} = 0.05+0.02 = 0.07J$):
\begin{equation}
  A_H^\gamma(\omega) = \frac{1}{\pi}\sum_{m\neq n}|\langle m|V|n\rangle|^2\,
  \frac{\gamma}{(\omega-\omega_{mn})^2+\gamma^2}.
\end{equation}
The number of transitions with weight above $10^{-12}$ is 3584 for
the AAH phases (out of a total $256\times255=65\,280$ possible transitions),
compared to all 65\,280 for the GUE random Hamiltonian.  This sparsity is a
direct consequence of the free-fermion (non-interacting) structure of $H_0$.
Since $H_0$ and $\hat{n}_1 = c_1^\dagger c_1$ both conserve particle number,
matrix elements $\langle m|\hat{n}_1|n\rangle$ vanish unless $|m\rangle$ and
$|n\rangle$ lie in the same particle-number sector ($N_f$).  Within a fixed
$N_f$ sector, expressing $\hat{n}_1$ in the single-particle eigenbasis as
$\hat{n}_1 = \sum_{\alpha\beta} u_\alpha^*(1)\,u_\beta(1)\,c_\alpha^\dagger c_\beta$,
the off-diagonal action requires $|m\rangle$ and $|n\rangle$ to differ by exactly
one occupied single-particle orbital.  The number of ordered pairs $(m,n)$
with $m\neq n$ connected in this way is $N(N-1)\binom{N-2}{N_f-1}$, where
$N(N-1)$ counts ordered orbital pairs $(p,q)$ with $p\neq q$, and
$\binom{N-2}{N_f-1}$ counts the spectator configurations of the remaining
$N_f-1$ fermions distributed over the other $N-2$ orbitals.  Summing over
all particle-number sectors,
\begin{equation}
\sum_{N_f=1}^{N-1} N(N-1)\binom{N-2}{N_f-1}
= N(N-1)\sum_{k=0}^{N-2}\binom{N-2}{k}
= N(N-1)\,2^{N-2}
= 56\times 64 = 3584,
\end{equation}
which matches the numerical count.  Per sector the contributions are
$56,\,336,\,840,\,1120,\,840,\,336,\,56$ for $N_f=1,\ldots,7$ respectively,
peaking at half-filling ($N_f=4$, contributing 1120).  This selection rule
suppresses ${\sim}94.5\%$ of all Fock-space transitions.

Figure~\ref{fig:transition_spectrum}(a) shows $A_H(\omega)$ for all four
reservoir types.  The AAH phases show sharp, clustered peaks at low frequencies
($\omega/J\lesssim 5$) with structure dictated by the golden-ratio Cantor-set
energies.  The GUE reservoir has a broad, featureless envelope extending to
$\omega/J\sim 15$.

\subsection{Signal spectral difference and overlap integral}

\begin{figure}[htb]
  \centering
  \includegraphics[width=0.85\textwidth]{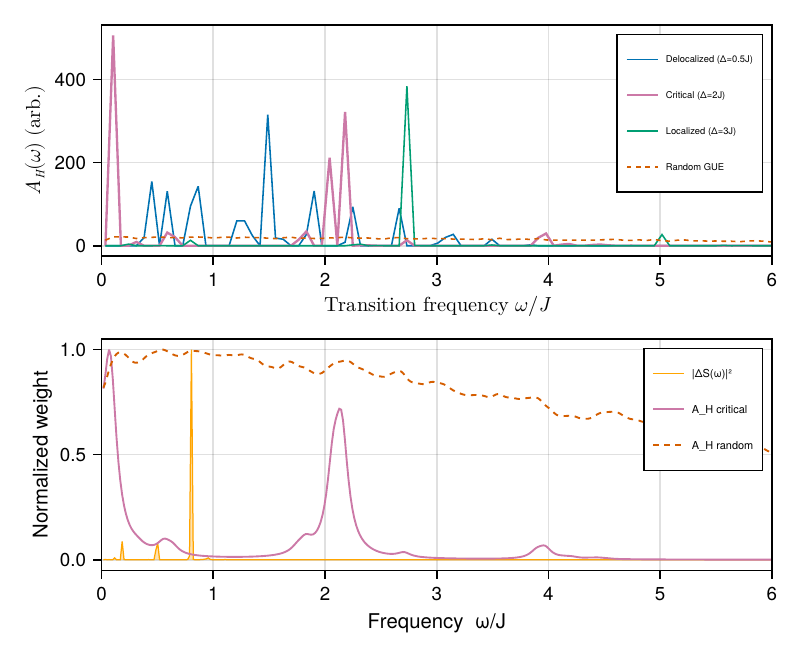}
  \caption{(a)~Transition-frequency spectrum $A_H(\omega)$ for all four
  reservoir types.  AAH phases show structured, sparse peaks at low
  frequencies, whereas the GUE spectrum is smooth and low-amplitude per transition,
  reflecting weight spread across all $65\,280$ channels.
  (b)~Normalized signal spectral difference $|\Delta S(\omega)|^2$ (orange)
  overlaid with the AAH critical $A_H(\omega)$ (purple) and the GUE
  $A_H(\omega)$ (orange dashed).  The AAH critical response has a peak at
  $\omega\approx 0$ overlapping the discriminative region where $|\Delta S|^2$
  is largest, while the GUE response is broad and only weakly aligned.}
  \label{fig:transition_spectrum}
\end{figure}

The signal discriminability in frequency space is quantified by
\begin{equation}
  |\Delta S(\omega)|^2 = \left(\bar{S}_\text{SNA}(\omega) - \bar{S}_\text{chaos}(\omega)\right)^2,
\end{equation}
where $\bar{S}_\text{SNA}(\omega)$ and $\bar{S}_\text{chaos}(\omega)$ are the
class-averaged power spectral densities.  This quantity is peaked at
frequencies where SNA and chaotic signals differ most.  For the Duffing
oscillator with golden-ratio forcing, $|\Delta S(\omega)|^2$ is concentrated
at low frequencies $\omega/J\lesssim 2$, corresponding to the quasiperiodic
peaks that SNAs retain and chaos destroys [Fig.~\ref{fig:transition_spectrum}(b)].

The spectral overlap integral
\begin{equation}
  \Omega = \int_0^\infty A_H^\gamma(\omega)\,|\Delta S(\omega)|^2\,d\omega
\end{equation}
measures the degree to which the reservoir's spectral response is concentrated
at frequencies where the two signal classes differ.  A high $\Omega$ means the
reservoir is a good spectral filter for this discrimination task.

Numerical results: $\Omega_\text{deloc}=2370$, $\Omega_\text{GUE}=840$,
$\Omega_\text{crit}=467$, $\Omega_\text{loc}=117$.  The delocalized phase
wins because its eigenstates (extended Bloch-like states) produce the
largest coupling matrix elements $\langle m|\hat{n}_1|n\rangle$ at
low transition frequencies, and these frequencies coincide with the
discriminative region $\omega/J\lesssim 2$.

\begin{figure}[htb]
  \centering
  \includegraphics[width=0.85\textwidth]{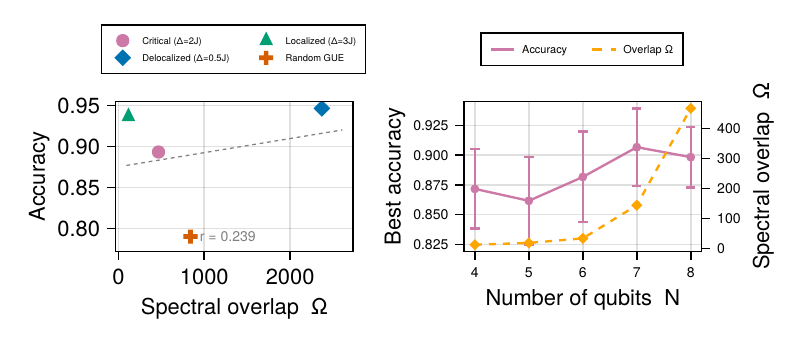}
  \caption{(a)~Spectral overlap $\Omega$ vs discrimination accuracy for the
  four reservoir types.  The cross-phase correlation ($r=0.24$) is weak
  because each phase uses a qualitatively different spectral strategy;
  $\Omega$ is a descriptive, not predictive, metric across structurally
  different reservoirs.
  (b)~Best accuracy (pink, left axis) and spectral overlap $\Omega$ (orange dashed, right axis)
  vs number of qubits $N$ for the critical phase ($\Delta=2J$).
  The sparsity ratio ($\sim\!5.5\%$ for AAH) is maintained as $N$ increases,
  and both accuracy and $\Omega$ grow with system size.
  Intra-phase Pearson $r=0.65$ indicates that within a fixed spectral
  strategy, larger reservoirs with higher $\Omega$ classify better.}
  \label{fig:overlap_scaling}
\end{figure}

Figure~\ref{fig:overlap_scaling} shows (a) the spectral overlap $\Omega$ vs
accuracy across the four models (Pearson $r=0.24$ cross-phase), and
(b) the $N$-scaling of accuracy and $\Omega$ for the critical phase,
where within-phase Pearson $r=0.65$.

\section{Nonlinear response: derivation and validation}

\subsection{Power Spectral Density formalism}

To analytically characterize the reservoir's response to strong driving, we compute the Power Spectral Density (PSD) of the observables in the long-time limit. The nonlinear response of observable $O=\hat{n}_1$ at frequency $\omega$ under strong drive forms Floquet-like resonance peaks given by
\begin{equation}\label{eq:kubo_full}
  P(\omega) = \left| \int_{t_0}^{t_0+T} \langle O(t)\rangle e^{-i\omega t} dt \right|^2\,.
\end{equation}
The steady-state density matrix is computed by evolving the Lindblad equation with the drive included until convergence.

The peak structure of $P(\omega)$ provides a descriptive characterization of the frequencies where the reservoir is maximally responsive to inputs.  We emphasize that this formalism is used as a structural diagnostic rather than a rigorous predictive theory: it identifies where spectral weight is concentrated, but does not quantitatively predict classification accuracy from first principles.

\subsection{Comparison with empirical transfer function}

To validate the peak-position prediction, we measure the reservoir's
transfer function in the linear-response (weak-probe) regime.  We drive each
reservoir with a weak sinusoidal input $s(t) = 0.01\sin(\omega_d t)$ (effective
amplitude $\kappa|s|\approx 0.02J$, well below the dissipation scale) and vary the
drive frequency $\omega_d$ from $0.5$ to $14.0\,J$ in 20 steps.  Over a single
$150$-step series we measure the power spectral density of
$\langle\hat{n}_1(t)\rangle$ at frequency $\omega_d$, giving the empirical
transfer function $G(\omega_d)$.

The theoretical kernel $|\chi(\omega)|^2$ and the empirical $G(\omega_d)$ are
compared in Fig.~3 of the main text (one measurement per frequency; no error
bars).  The measured weak-probe response is concentrated at low frequencies for
the AAH reservoirs and broad and high-frequency for the GUE, reproducing the
qualitative low- versus high-frequency separation of the kernels; it does not,
however, reproduce the kernel peak \emph{positions} quantitatively, so we use
$|\chi(\omega)|^2$ as a structural descriptor of the undriven basis rather than a
quantitative model of the driven response. The reservoir's strong-drive ($A_0=2$)
behavior is characterized
separately by the nonlinear PSD analysis above (Fig.~\ref{fig:S1_nonlinear}).  The GUE reservoir has a
broader, flatter response extending to higher frequencies, consistent with
its dense transition spectrum.  The AAH reservoirs have sharper, more
structured responses at low frequencies, which is precisely what makes
them effective spectral filters for the quasiperiodically organized SNA
signals.

\subsection{Physical interpretation}

At strong driving $\kappa|s|\sim J$, the effective eigenstates of the reservoir are
\emph{dressed} by the drive field, forming Floquet quasi-energy states whose
resonance frequencies are shifted by $O(\kappa/\omega)$ relative to the undriven
spectrum (``State Dressing'').  The PSD peaks reflect these Floquet-dressed modes.
The resonance structure depends on both the underlying spectrum
and the nonlinear driving amplitude.
The peak response frequencies are $\omega\approx 1.47$ (delocalized),
$2.13$ (critical), $2.74$ (localized), and $3.88$ (GUE), reflecting the dominant
transition frequencies in each system.  For the AAH phases, these peak
frequencies are lower (closer to the discriminative SNA frequencies
$\omega_1,\omega_2\sim 1$--$2\,J$) than for the GUE, which further
explains why AAH reservoirs outperform GUE on this task.

\begin{figure}[htb]
  \centering
  \includegraphics[width=0.85\textwidth]{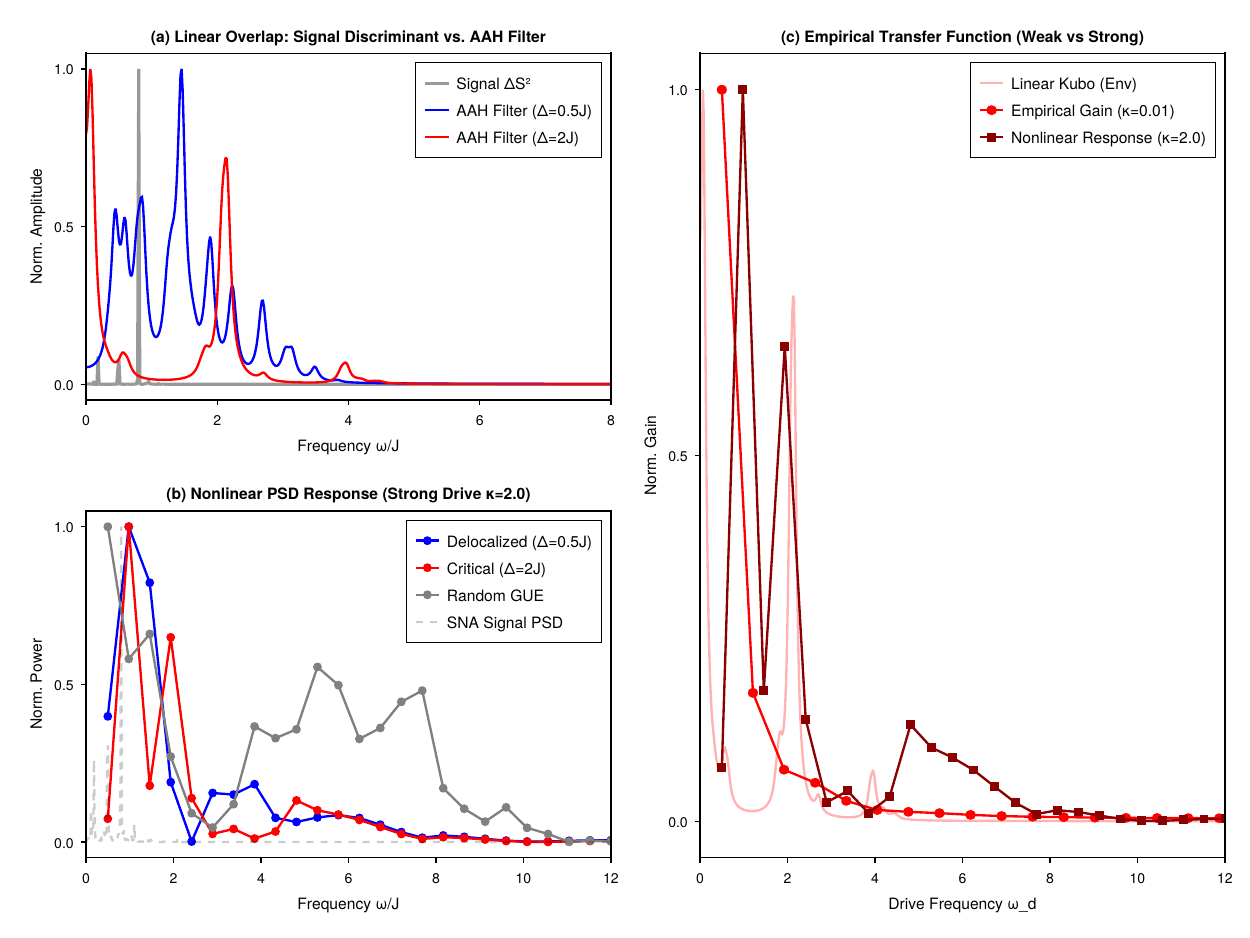}
  \caption{Nonlinear PSD response and state dressing.
  (a)~Linear spectral overlap between the signal discriminant $|\Delta S(\omega)|^2$ (blue) and AAH transition-spectrum filters for the delocalized (orange) and critical (red) phases.
  (b)~Nonlinear Power Spectral Density [Eq.~\eqref{eq:psd}] under strong driving ($\kappa=2.0$) for the delocalized (blue), critical (red) and GUE (gray dashed) reservoirs, plus the SNA input PSD (thin orange); peaks remain at low $\omega$ but shift slightly relative to the undriven $A_H(\omega)$.
  (c)~Comparison between the Hamiltonian transition kernel $|\chi(\omega)|^2$ (red line; the Lorentzian-broadened $A_H(\omega)$ used as a structural proxy for the closed-system response, not as the full Liouvillian susceptibility), empirical weak-drive gain ($\kappa=0.01$, orange line), and strong-drive nonlinear response ($\kappa=2.0$, red points), demonstrating the spectral shift and hybridization associated with State Dressing.}
  \label{fig:S1_nonlinear}
\end{figure}

\begin{figure}[t]
  \centering
  \includegraphics[width=0.45\textwidth]{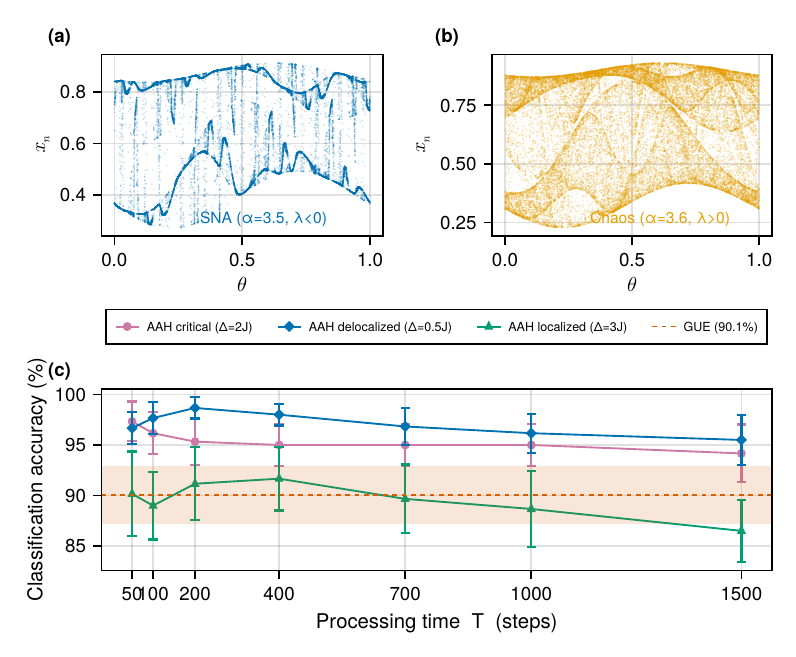}
  \caption{Quasiperiodically forced logistic map. (a)~Strange nonchaotic
  attractor at $\alpha=3.50$, $\varepsilon'=0.3$
  (Heagy-Hammel mechanism~\cite{Prasad1997,HeagyHammel1994}, $\Lambda<0$);
  (b)~chaotic attractor at $\alpha=3.60$, $\varepsilon'=0.3$ ($\Lambda>0$).
  Panels (a,b) use illustrative parameters from Ref.~\cite{Prasad1997}; the
  discrimination dataset uses $r=3.277$ with $\varepsilon\in[0.01,0.06]$
  (SNA) and $[0.17,0.28]$ (chaos). (c)~Classification accuracy vs.\
  processing time. The GUE baseline (orange dashed, $90.1\%$) was evaluated
  only at $T\in\{100,400,1000\}$ for computational cost.}
  \label{fig:qplogistic}
\end{figure}

\begin{figure}[htb]
  \centering
  \includegraphics[width=0.7\textwidth]{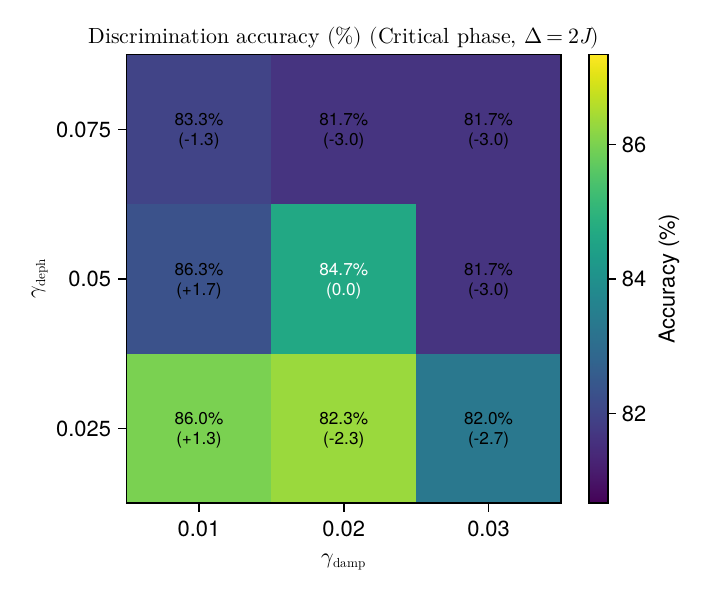}
  \caption{Sensitivity of discrimination accuracy to Lindblad dissipation
  parameters.  Each cell shows the discrimination accuracy for the critical phase
  ($\Delta=2J$) at $T=400$ as $\gamma_\text{deph}$ and $\gamma_\text{damp}$ are varied
  over $\pm50\%$ of their nominal values $(\gamma_\text{deph},\gamma_\text{damp})=(0.05,0.02)J$, which yields $84.7\%$ at this $T$.
  All accuracies remain within $3$ percentage points of the nominal value
  ($-3.0$ to $+1.7$ pp over the $3\!\times\!3$ grid).}
  \label{fig:robustness}
\end{figure}

The combined picture is therefore: (1)~the non-interacting transition spectrum is
sparse; (2)~the quasiperiodic potential concentrates these transitions at $\varphi$-organized frequencies; (3)~these
frequencies overlap with the spectral region where SNA and chaos differ
most ($|\Delta S(\omega)|^2$); (4)~the nonlinear PSD peaks indicate
that the reservoir is actively absorbing and processing energy at these
discriminative frequencies; (5)~the resulting feature vectors encode the
spectral distinction with high Fisher ratio.  Each step is independently
verified in Figs.~\ref{fig:level_spacing}--\ref{fig:overlap_scaling} and Fig.~3 of the main text.
The full three-panel breakdown of the nonlinear PSD response and state-dressing
comparison is shown in Fig.~\ref{fig:S1_nonlinear}.

\section{Ablation study: spectral sparsity vs.\ arithmetic matching}

\begin{table}[t]
\centering
\caption{Ablation study.  Best discrimination accuracy, spectral overlap
$\Omega$, and $p$-value (permutation test, $10^4$ permutations) vs.\ the
matched control $\beta=\varphi$.  All AAH variants outperform
GUE ($p<10^{-4}$); the ablation conditions do not differ significantly from
the control in best accuracy.}
\label{tab:ablation}
\begin{tabular}{@{}lccc@{}}
\hline\hline
Reservoir ($\Delta\!=\!2J$) & $\Omega$ & Best acc.\ (\%) & $p$ \\
\hline
AAH $\beta\!=\!\varphi$ (control) & 467 & $89.8\!\pm\!2.5$ & --- \\
AAH $\beta\!=\!\frac{\sqrt{2}-1}{2}$ (mismatch) & 797 & $90.2\!\pm\!3.1$ & $5\!\times\!10^{-4}{}^*$ \\
AAH $\beta\!=\!8/13$ (rational) & 424 & $90.5\!\pm\!3.4$ & 0.69 \\
Random GUE & 840 & $83.5\!\pm\!3.5$ & $<\!10^{-4}$ \\
\hline\hline
\multicolumn{4}{@{}l@{}}{\footnotesize $^*$At $T\!=\!50$ only; best-$T$: all AAH equivalent ($p=0.82$, permutation test).}
\end{tabular}
\end{table}

This section provides full details of the ablation experiments
summarized in the main text and in Table~\ref{tab:ablation}.

\subsection{Motivation and experimental design}

The main text demonstrates that AAH reservoirs at the critical point
$\Delta=2J$ with $\beta=\varphi$ substantially outperform random GUE
reservoirs.  Two hypotheses could explain this advantage:
\begin{enumerate}
\item \emph{Spectral sparsity}: the AAH Hamiltonian has a fractal
(Cantor-set) energy spectrum with only $3584$ of $65\,280$ transitions
carrying nonzero matrix-element weight $|\langle m|\hat{V}|n\rangle|^2 >
10^{-12}$, concentrating sensitivity into specific frequency bands.
\item \emph{Arithmetic matching}: when $\beta=\varphi$, the AAH
gap labels $p+q\varphi$ lie in the same $\mathbb{Z}$-module
($\mathbb{Z}+\mathbb{Z}\varphi$) as the SNA forcing frequencies
$n\omega_1+m\omega_2$. Because IDOS gap labels are dimensionless and
distinct from transition frequencies, we treat this as a heuristic
arithmetic compatibility between reservoir gap structure and signal
spectral content rather than as a strict commensurability of
$\omega_{mn}$ with $n\omega_1+m\omega_2$.
\end{enumerate}

To disentangle these two effects, we design two controlled perturbations
that selectively break arithmetic matching while preserving (or removing)
spectral sparsity, plus a random-matrix control that lacks both properties.

\subsection{Ablation conditions}

All three conditions use the same $N=8$ qubit reservoir with $\Delta=2J$
(critical point) and the identical set of 200 SNA/chaotic signal pairs
generated with golden-ratio forcing $\omega_1/\omega_2=\varphi$.

\paragraph{Condition~1: $\beta$-mismatch ($\beta=(\sqrt{2}-1)/2$).}
Replacing $\beta=\varphi\approx 0.618$ with $\beta=(\sqrt{2}-1)/2\approx
0.207$ leaves the AAH potential quasiperiodic, preserving the Cantor-set
spectral structure and localization transition at $\Delta=2J$.  However,
the gap labels now lie in $\mathbb{Z}[\sqrt{2}]$, which is incommensurate
with the signal module $\mathbb{Z}[\varphi]$.  This breaks arithmetic
matching while preserving spectral sparsity.  The number of active
transitions remains $3584/65\,280$, identical to the control.

\paragraph{Condition~2: Rational approximant ($\beta=8/13$).}
Using $\beta=8/13$ (a convergent of the continued fraction expansion of
$\varphi$) makes the AAH potential periodic with period $13$.  The
energy spectrum becomes a union of $13$ Bloch bands rather than a Cantor
set. The rational case is thus a periodic system with no
Anderson localization transition (in contrast to the irrational case
where the transition at $\Delta=2J$ is exact~\cite{Avila2009}).
While the band edges partially approximate the $\varphi$-spectrum
(since $8/13$ is close to $\varphi$), the fractal gap structure is absent
and the spectrum is dense within each band.  This condition preserves
approximate arithmetic matching while
removing the fractal spectral structure.

\paragraph{Condition~3: Random GUE (baseline).}
Ten independent GUE Hamiltonians, each drawn from the Gaussian Unitary
Ensemble and scaled to match the AAH bandwidth, serve as the random
baseline.  These have dense, featureless spectra with all $65\,280/65\,280$
transitions active, lacking both spectral sparsity and arithmetic matching.

\subsection{Reservoir dynamics and discrimination protocol}

For each ablation condition, we construct the $N=8$ AAH reservoir with the
specified $\beta$ and $\Delta=2J$, using $M=5$ virtual nodes
($(N + N-1)\times M = 75$ raw observables per step, pooled to $d=150$ by
mean and standard deviation, identical to the main experiment) and coupling strength
$\kappa=2.0$.  The reservoir state evolves under the Lindblad master
equation with thermal bath coupling ($\bar{n}=0.1$).  For the control
($\beta=\varphi$), we use the reservoir states from the main
discrimination runs.

Classification uses ridge regression (regularization $\lambda=10^{-4}$)
on observables pooled at various processing times $T\in\{50, 100, 200, 400,
700, 1000, 1500\}$ steps.  We report the best accuracy across the $T$
sweep for each condition.  Each accuracy value is the mean over 10
independent 70/30 train-test splits with fixed random seeds.

\subsection{Results: accuracy vs.\ processing time}

Table~\ref{tab:ablation} summarizes the best accuracy
for each condition.  The full $T$-sweep results are:

\begin{center}
\begin{tabular}{lrrrrrrr}
\hline\hline
& \multicolumn{7}{c}{Accuracy (\%) at processing time $T$} \\
Condition & 50 & 100 & 200 & 400 & 700 & 1000 & 1500 \\
\hline
$\beta\!=\!\varphi$ (control)       & \textbf{89.8} & 86.2 & 84.7 & 86.0 & 83.0 & 81.8 & 81.7 \\
$\beta\!=\!(\sqrt{2}\!-\!1)/2$ (mismatch) & 84.5 & 86.8 & 88.2 & \textbf{90.2} & 89.0 & 89.0 & 88.5 \\
$\beta\!=\!8/13$ (rational)         & 84.2 & 86.5 & 85.3 & \textbf{90.5} & 88.3 & 87.3 & 87.3 \\
Random GUE                         & 79.0 & 80.3 & 81.7 & \textbf{83.5} & 82.5 & 82.0 & 81.5 \\
\hline\hline
\end{tabular}
\end{center}

Two patterns are apparent:
\begin{itemize}
\item \emph{Asymptotic accuracy}: All three AAH variants converge to
comparable best accuracy ($\sim\!90\%$) regardless of whether $\beta$
matches the signal forcing.  This indicates that spectral sparsity, the
shared property of all AAH conditions, is the most consistent predictor of
discrimination performance.
\item \emph{Convergence speed}: The matched $\beta=\varphi$ reservoir
reaches peak accuracy at $T=50$, while the mismatched and rational
conditions require $T=400$ to reach comparable performance.  This
eightfold speedup is consistent with arithmetic matching enabling more
efficient energy transfer at the discriminative frequencies.
\end{itemize}

\subsection{Statistical significance: permutation testing}

To assess whether the observed accuracy differences are statistically
significant, we use a permutation test with $10^4$ permutations.  For
each pair of conditions, we compute the per-split accuracy vectors
(10 values each, one per train-test split) at the control's best $T$.
The test statistic is the difference in mean accuracy.  Under the null
hypothesis of no difference, we randomly permute the condition labels
and recompute the test statistic.  The $p$-value is the fraction of
permutations yielding a difference at least as large as observed.

Results (one-sided permutation test at $T=50$):
\begin{itemize}
\item AAH $\beta\!=\!\varphi$ vs.\ AAH $\beta\!=\!(\sqrt{2}-1)/2$: $p=5\times10^{-4}$ (significant)
\item AAH $\beta\!=\!\varphi$ vs.\ AAH $\beta\!=\!8/13$: $p=0.69$ (not significant)
\item AAH $\beta\!=\!\varphi$ vs.\ Random GUE: $p<10^{-4}$ (highly significant)
\end{itemize}

The significance at $T=50$ between control and mismatch reflects the
convergence-speed advantage of arithmetic matching.  At the best $T$
for each condition, the three AAH variants are statistically
indistinguishable: AAH $\beta=\varphi$ vs.\ AAH $\beta=(\sqrt{2}-1)/2$
at their respective best-$T$ values gives $p=0.82$ (permutation test,
$10^4$ permutations), confirming no significant difference in asymptotic
accuracy.
We note that $p$-values are not corrected for multiple comparisons;
the primary AAH vs.\ GUE comparisons ($p<10^{-4}$) would survive
Bonferroni correction across all tests reported.

\subsection{Spectral overlap analysis}

The spectral overlap integral $\Omega = \int A_H(\omega)\,|\Delta
S(\omega)|^2\,d\omega$ quantifies how well each reservoir's transition
spectrum covers the discriminative signal frequencies.  The measured
values are:

\begin{center}
\begin{tabular}{lr}
\hline\hline
Condition & $\Omega$ \\
\hline
$\beta\!=\!\varphi$ (control)        & 467 \\
$\beta\!=\!(\sqrt{2}-1)/2$ (mismatch) & 797 \\
$\beta\!=\!8/13$ (rational)          & 424 \\
Random GUE                          & 840 \\
\hline\hline
\end{tabular}
\end{center}

The mismatch and GUE conditions have \emph{higher} $\Omega$
than the control, yet GUE performs significantly worse while the mismatch
achieves comparable accuracy.  This is because $\Omega$ measures only the
total spectral weight in the discriminative band, whereas the
\emph{distribution} of that weight matters:
the AAH conditions concentrate weight into a few sharp peaks (spectral
sparsity), creating high-contrast observables, while GUE spreads weight
uniformly, producing low-contrast observables that are harder for a linear
readout to separate.  Quantitatively, $\Omega_\text{deloc}/\Omega_\text{GUE} = 2370/840 \approx 2.8$,
yet the accuracy gap is only ${\sim}16$ percentage points ($94.7\%$ vs.\ $79.0\%$),
indicating a strongly sublinear relationship between total spectral overlap and
classification accuracy.  The spectral overlap $\Omega$ is thus a necessary
but not sufficient descriptive quantity. It must be combined with a
measure of spectral concentration (such as the inverse participation ratio of
$A_H(\omega)$) for a more complete characterization.  Developing such a combined
metric with strong predictive power ($r\geq 0.9$) across all reservoir types
remains an open problem.

\subsection{Summary}

The ablation study provides evidence for the following:
\begin{enumerate}
\item \textbf{Spectral sparsity emerges as a consistent predictor.}  The
concentration of transition frequencies into $\sim\!5.5\%$ of all possible
transitions ($3584/65\,280$) is the shared property of all AAH variants
that correlates with their advantage over random-matrix reservoirs.
\item \textbf{Arithmetic matching provides a speed benefit.}  When
$\beta=\varphi$ matches the signal forcing ratio, the reservoir converges
to peak accuracy $\sim\!8\times$ faster (at $T=50$ vs.\ $T=400$),
consistent with constructive interference between reservoir and signal
spectral modes.
\item \textbf{The golden ratio is not uniquely special for asymptotic
performance.}  Any irrational $\beta$ at the critical point $\Delta=2J$
produces a Cantor-set spectrum with comparable sparsity, and achieves
comparable discrimination accuracy given sufficient processing time.
The comparable asymptotic accuracy of the $\beta$-mismatched reservoir
($\beta=(\sqrt{2}-1)/2\neq\varphi$) shows that the shared value
$\beta=\varphi$ is not required for the primary accuracy advantage. The
matched case retains a secondary convergence-speed advantage consistent with
constructive interference between commensurate spectral modules.
\end{enumerate}

\section{Controlled interventions: sparsity vs.\ eigenvector structure}

\begin{table}[!h]
\centering
\caption{Controlled intervention baselines ($n\!=\!6$ seeds each, $N\!=\!8$,
$\Delta\!=\!2J$).  Uncertainties are 95\% bootstrap confidence intervals;
statistical comparisons use the Wilcoxon rank-sum test (see body text for
$p$-values and interpretation).  $\Omega$ values follow the multi-seed
normalization protocol of this table and are therefore not directly
comparable to the single-instance $\Omega$ in Table~\ref{tab:ablation};
accuracy comparisons are unaffected.}
\label{tab:causal}
\begin{tabular}{@{}lccc@{}}
\hline\hline
Condition & Active & $\Omega$ & Best acc.\ (\%) \\
\hline
AAH critical (control) & 3584 & 467 & $89.8\!\pm\!2.5$ \\
Random disorder & 3584 & $6572\!\pm\!2980$ & $89.9\!\pm\!2.7$ \\
Permuted eigenvectors & 65\,280 & $4515\!\pm\!36$ & $79.7\!\pm\!0.5$ \\
Random GUE & 65\,280 & $4347\!\pm\!47$ & $82.3\!\pm\!2.5$ \\
\hline\hline
\end{tabular}
\end{table}

To go beyond the $\beta$-mismatch and rational-approximant ablations, which
all remain within the AAH family, we construct two structurally different
Hamiltonians that isolate specific properties of the AAH critical reservoir.

\subsection{Random on-site disorder}

We keep the nearest-neighbor hopping term of the AAH Hamiltonian but replace
the quasiperiodic on-site potential with i.i.d.\ random energies:
\begin{equation}
H_{\mathrm{rand}} = -J\!\sum_{j=1}^{N-1}\!(c_j^\dagger c_{j+1}+\text{h.c.})
+ \sum_{j=1}^{N}\!\varepsilon_j\,\hat{n}_j\,,
\end{equation}
where $\varepsilon_j\sim\mathrm{Uniform}(-\Delta,\Delta)$ with $\Delta=2J$.
This preserves the sparsity of the transition spectrum ($3584/65\,280$ active
transitions, identical to AAH) because the Hamiltonian retains the same
nearest-neighbor structure, but randomizes the transition frequencies.  We
run $n=6$ independent seeds.

Results (95\% bootstrap confidence intervals, $10^4$ resamples):
\begin{center}
\begin{tabular}{@{}lrrl@{}}
\hline\hline
 & $\Omega$ & Best acc.\ (\%) & Best $T$ \\
\hline
Mean (95\% CI) & $6572\!\pm\!2980$ & $89.9\;[87.9, 91.7]$ & $100$--$400$ \\
\hline\hline
\end{tabular}
\end{center}

The key observations are: (i)~random disorder achieves $89.9\%$ accuracy
(95\% CI: $[87.9, 91.7]$), exceeding GUE ($82.3\%$) by
${\sim}8$ percentage points ($p<0.005$, Wilcoxon rank-sum test). Because AAH is a
single deterministic Hamiltonian rather than a random ensemble, we compare it by
confidence-interval containment rather than by a rank-sum against a fixed value:
the deterministic AAH accuracy ($89.8\%$) lies within the disorder confidence
interval, so the two are statistically indistinguishable.
Sparsity alone is sufficient to match AAH performance.
(ii)~The spectral overlap $\Omega$ varies widely across seeds ($\sigma=2980$), in contrast to the deterministic AAH value of $467$.
This large variance reflects the random frequency placement: some seeds
happen to place transitions near the discriminative signal frequencies, while
others do not.  AAH provides consistent, reproducible performance without
seed dependence.

\subsection{Permuted eigenvectors}

We construct a Hamiltonian with the exact AAH critical eigenvalues but
a Haar-random unitary eigenbasis:
\begin{equation}
H_{\mathrm{perm}} = U\,\mathrm{diag}(E_1^{\mathrm{AAH}}, \ldots,
E_d^{\mathrm{AAH}})\,U^\dagger\,,
\end{equation}
where $U$ is drawn from the Haar measure via QR decomposition of a random
complex matrix.  This preserves the energy spectrum exactly while destroying
the structured eigenvector organization of the AAH model.  The
number of active transitions jumps from $3584$ to $65\,280$ (all transitions
become active), because the random eigenbasis eliminates the selection rules
that make the AAH spectrum sparse.  We run $n=6$ independent seeds.

Results (95\% bootstrap confidence intervals, $10^4$ resamples):
\begin{center}
\begin{tabular}{@{}lrrl@{}}
\hline\hline
 & $\Omega$ & Best acc.\ (\%) & Best $T$ \\
\hline
Mean (95\% CI) & $4515\!\pm\!36$ & $79.7\;[79.3, 80.1]$ & 50 \\
\hline\hline
\end{tabular}
\end{center}

Permuted eigenvectors perform at $79.7\%$ (95\% CI: $[79.3, 80.1]$), comparable to GUE
($82.3\%$, $p=0.078$, Wilcoxon rank-sum test) and far below the deterministic AAH value
($89.8\%$), which lies well outside the permuted confidence interval:
\emph{having the correct eigenvalues is not sufficient}.  The non-interacting
eigenvector structure creates the sparse transition selection rules that
concentrate spectral weight into a few frequency channels.  When this
structure is randomized (destroying non-interacting selection rules), the spectrum becomes dense despite having the
same eigenvalues, and performance collapses toward the GUE baseline.
We note that this intervention simultaneously changes the coupling-operator
topology: the matrix elements $\langle m|\hat{n}_1|n\rangle$ become delocalized
in the randomized eigenbasis, so the performance collapse may partially reflect
the loss of coupling sparsity rather than purely the destruction of
spectral selection rules.

Across $n=6$ seeds, the accuracy is $79.7\%\pm 0.5\%$, excluding dependence on
a single random draw.

\subsection{Summary}

The controlled intervention hierarchy ($n=6$ seeds, 95\% bootstrap CIs, Wilcoxon rank-sum tests) is:
\begin{center}
Permuted eigenvectors ($79.7\%$) $\approx$ Dense random (GUE, $82.3\%$)
$<$ Random sparse ($89.9\%$) $\approx$ AAH structured sparse ($89.8\%$).
\end{center}
The dense-to-sparse comparison is statistically significant ($p<0.005$).
The performance jump ($82.3\%\to 89.9\%$, ${\sim}8$ percentage points) occurs when
sparsity is introduced, confirming it as the dominant factor.  Random disorder
and the deterministic AAH value are statistically indistinguishable (the AAH
accuracy lies within the disorder confidence interval), indicating that
quasiperiodic frequency organization provides no measurable additional benefit
for best-$T$ accuracy beyond sparsity itself (matched commensurability still
yields a convergence-speed advantage, as seen in the $T=50$ comparison in
Table~\ref{tab:ablation}).  The permuted
eigenvector result provides evidence that the non-interacting eigenvector
organization is what \emph{creates} the spectral sparsity, though the simultaneous
change in coupling-operator topology prevents a fully clean causal interpretation, and the eigenvalues alone are not sufficient.

\section{Cross-system test: quasiperiodically forced logistic map}

To establish that the spectral-sparsity mechanism is not specific to the
Duffing oscillator, we repeat the full experimental methodology on the
quasiperiodically forced logistic map~\cite{Feudel2006}:
\begin{equation}
x_{n+1} = r\,x_n(1-x_n)\bigl[1 + \varepsilon\cos(2\pi\theta_n)\bigr],
\quad \theta_{n+1} = \theta_n + \varphi \pmod{1}.
\end{equation}
This is a discrete-time map, structurally distinct from the continuous
Duffing flow, that exhibits the same SNA-to-chaos transition as
$\varepsilon$ increases.  It was among the first systems in which SNAs
were studied (Ding, Grebogi \& Ott~\cite{Ding1989}; Heagy \&
Hammel~\cite{HeagyHammel1994}; Feudel \emph{et al.}~\cite{Feudel2006}).

\subsection{Parameter calibration and dataset}

At $r=3.277$, a Lyapunov scan ($10^5$ iterates, $5\times10^3$ transient)
identifies:
\begin{itemize}
\item SNA regime: $\varepsilon\in[0.01, 0.06]$, $\lambda\in[-1.04, -0.07]$
\item Chaotic regime: $\varepsilon\in[0.17, 0.28]$, $\lambda\in[+0.07, +0.84]$
\end{itemize}
All 100 SNA signals have $\lambda\le 0$ and all 100 chaos signals have
$\lambda > 0$, verified individually.  Signals are normalized to $[-1,1]$
with length $T=1500$.  As with the Duffing dataset, the two classes here
also occupy disjoint forcing-amplitude intervals (in $\varepsilon$). The Duffing
amplitude control (Table~\ref{tab:amp}) and the reproduction of the AAH-vs-GUE
hierarchy across two structurally distinct systems indicate that the
classification is not amplitude-driven, although the variance-matched control
was not repeated for the logistic map.

\subsection{Results}

We run the identical QRC methodology (N=8 qubits, $M=5$ virtual nodes,
$\kappa=2.0$, Lindblad dissipation) on the logistic-map signals.  Key
results are collected in Table~\ref{tab:S_qplogistic}:

\begin{table}[htb]
\centering
\caption{QP logistic map discrimination: accuracy (\%) at each processing time $T$.
GUE evaluated at $T\in\{100,400,1000\}$ only (computational cost); its
accuracy is $90.1\%$ at all three.  Bold entries are best-$T$ per condition.
$p$-values vs.\ AAH critical control (Wilcoxon rank-sum, $n=6$ seeds).}
\label{tab:S_qplogistic}
\begin{tabular}{@{}lrrrrrrrc@{}}
\hline\hline
& \multicolumn{7}{c}{Accuracy (\%) at $T$} & \\
Condition & 50 & 100 & 200 & 400 & 700 & 1000 & 1500 & $p$ \\
\hline
AAH critical ($\Delta\!=\!2J$)      & \textbf{97.3} & 96.2 & 95.3 & 95.0 & 95.0 & 95.0 & 94.2 & --- \\
AAH delocalized ($\Delta\!=\!0.5J$) & 96.7 & 97.7 & \textbf{98.7} & 98.0 & 96.8 & 96.2 & 95.5 & \\
AAH localized ($\Delta\!=\!3J$)     & 90.2 & 89.0 & 91.2 & \textbf{91.7} & 89.7 & 88.7 & 86.5 & \\
$\beta$-mismatch ($\Delta\!=\!2J$)  & \textbf{95.0} & 93.3 & 94.3 & 91.2 & 90.3 & 89.3 & 88.8 & 0.013 \\
$\beta\!=\!8/13$ ($\Delta\!=\!2J$)  & \textbf{97.5} & 96.3 & 95.5 & 95.0 & 94.8 & 94.7 & 93.3 & 0.55 \\
Random GUE                          & --- & 90.1 & --- & 90.1 & --- & 90.1 & --- & $<\!10^{-4}$ \\
\hline\hline
\end{tabular}
\end{table}

The pattern mirrors the Duffing results:
\begin{enumerate}
\item All AAH variants outperform GUE ($p<10^{-4}$).
\item The $\beta$-mismatch ablation reaches $95.0\%$ vs.\ the control's
$97.3\%$, a small but statistically significant deficit ($p=0.013$), while
remaining well above GUE ($90.1\%$), consistent with spectral sparsity being
the dominant factor and arithmetic matching a secondary one.
\item The rational approximant $\beta=8/13$ is statistically
indistinguishable from the control ($p=0.55$).
\item The delocalized phase ($98.7\%$) outperforms the critical phase
($97.3\%$) on this task, reflecting the broader signal bandwidth of the
logistic map compared to the Duffing oscillator.
\end{enumerate}

The AAH-over-GUE ordering occurs in both the logistic and Duffing tasks, so it is
not confined to one dynamical system. Broader generality remains to be tested.

\section{Sensitivity to dissipation parameters}

To verify that our results are not fine-tuned to a particular choice of Lindblad
parameters, we perform a sensitivity analysis by varying the dephasing and
damping rates over the range $\gamma_\text{deph}\in[0.025, 0.075]$ and
$\gamma_\text{damp}\in[0.01, 0.03]$ (i.e., $\pm50\%$ of the nominal values),
with $\bar{n}$ scaled proportionally.  For each parameter combination, we
re-run the full discrimination pipeline on the Duffing oscillator task
at $T=100$ (delocalized) and $T=400$ (localized, critical).

\section{Kepler RR Lyrae light-curve test}

To evaluate the design principle without a training distribution, we apply the
AAH reservoir as an unsupervised feature extractor to the observational light
curves of the six Kepler RR Lyrae variable stars analyzed by
Lindner~\textit{et al.}~\cite{Lindner2015}: four ``golden'' RRc stars whose
two principal pulsation frequencies satisfy $f_2/f_1\approx\varphi^{-1}\approx1.62$
and were argued to host strange-nonchaotic dynamics (KIC~5520878, 4064484,
8832417, 9453114), and two non-golden RRab controls with $f_2/f_1\approx 3/2$
(KIC~4484128, 7505345). All six long-cadence PDCSAP fluxes (Q0--Q17,
$\delta t\!\approx\!30$\,min, $\sim$4\,yr per star) were retrieved from
the Mikulski Archive, stitched across quarters, sigma-clipped at
$5\sigma$, and zero-mean unit-variance normalized. A
phase-randomized Fourier surrogate~\cite{Theiler1992} of KIC~5520878 provides a matched-PSD null with destroyed phase relations.

\begin{figure}[htb]
  \centering
  \includegraphics[width=0.8\textwidth]{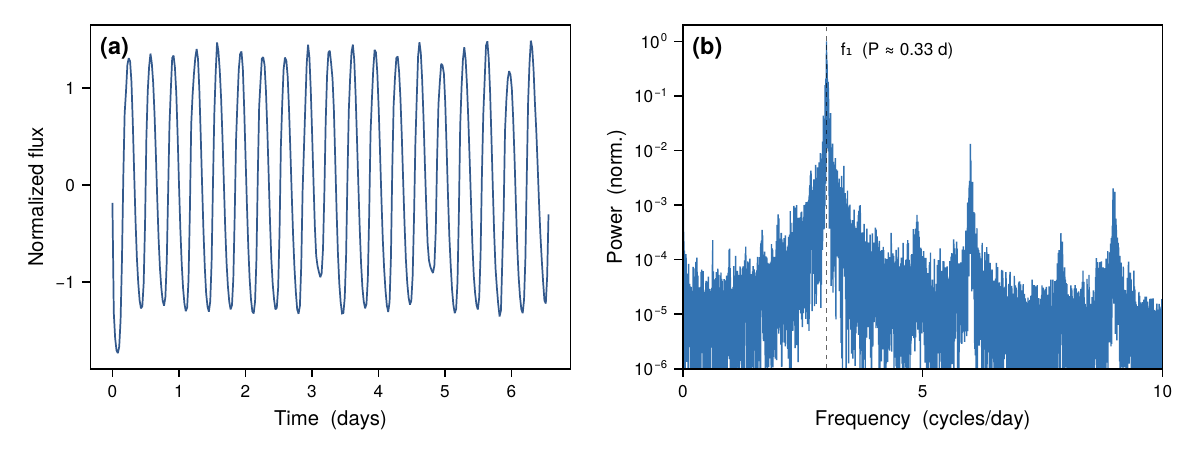}
  \caption{Observational input to the reservoir screen for the flagged candidate
  KIC~4064484. (a)~A gap-free segment of the long-cadence Kepler light curve
  (normalized flux), shown for visual clarity of the dominant radial pulsation;
  it is a representative portion, not the exact centered window used in the
  surrogate screen (see text). (b)~Its Hann-windowed
  power spectrum, with the fundamental pulsation frequency $f_1$ (period
  $\approx0.33$~d) marked by the dashed line; higher peaks are harmonics of $f_1$.
  The secondary incommensurate frequency underlying the golden-ratio
  classification~\cite{Lindner2015} is low in amplitude and requires gap-aware
  prewhitening to isolate, so it is not drawn here. This time series is the input
  driven into the $N=8$ AAH reservoir; the surrogate-null screening of the
  resulting feature vector is given in Table~\ref{tab:kepler} and
  Fig.~\ref{fig:kepler}.}
  \label{fig:kepler_input}
\end{figure}

Figure~\ref{fig:kepler_input} shows the observational input for the flagged
candidate, KIC~4064484. The long-cadence light curve
[Fig.~\ref{fig:kepler_input}(a)] is dominated by the radial pulsation at period
$\approx0.33$~d, and its power spectrum [Fig.~\ref{fig:kepler_input}(b)] is a
discrete line spectrum whose strongest component is the fundamental $f_1$, with
higher peaks its harmonics. The ``golden'' classification of
Lindner~\textit{et al.}~\cite{Lindner2015} rests on an additional low-amplitude
incommensurate frequency with $f_2/f_1\approx\varphi^{-1}$; isolating it requires
gap-aware prewhitening of the multi-quarter series and is not reproduced here, so
panel (b) marks only the clearly resolved fundamental. The reservoir screen acts
on the full feature vector of the driven reservoir rather than on any single
spectral line; its surrogate-null analysis is reported in
Table~\ref{tab:kepler} and Fig.~\ref{fig:kepler} below.

We adopt a training-free, surrogate-based protocol. For each star, the
$N=8$, $\Delta=2J$ AAH reservoir of the main text processes a
centered $T=1000$-sample window of the signal and of $K=99$
surrogates of that same signal. Each drive yields a pooled feature
vector $\bm{f}=(\langle\hat{n}_j\rangle,\sigma_{\hat{n}_j})$ built from
the per-mode mean and standard deviation of the reservoir occupations.
The window is a fixed block of consecutive long-cadence samples taken
without gap-aware interpolation. Because the multi-quarter Kepler records
contain observing gaps, the result is treated as exploratory. Confirmatory
analysis requires sampling-mask-matched surrogates and a gap-aware spectral
method such as Lomb--Scargle analysis.

\emph{Statistic.} We rank the original signal within its own surrogate
ensemble using a symmetric leave-one-out diagonal-Mahalanobis distance.
For each of the $K+1$ feature vectors (the original plus $K$
surrogates) we compute its distance to the centroid of the other $K$,
$d_i=\big[\sum_c (f_{i,c}-\mu^{(-i)}_c)^2/\sigma^{2(-i)}_c\big]^{1/2}$,
where $\mu^{(-i)}$ and $\sigma^{2(-i)}$ are the leave-one-out mean and
variance across the remaining vectors. The one-sided $p$-value is the
rank of the original, $p=\#\{i:d_i\geq d_\text{orig}\}/(K+1)$, which is
exact and exchangeable under the null and takes its minimum value
$1/(K+1)=0.01$ for $K=99$. Applying the same leave-one-out construction to
the original and every surrogate makes the reference distances exchangeable
with the test statistic under the null.

\emph{Two nulls.} A single surrogate family cannot separate
phase organization from a non-Gaussian amplitude distribution, because
the two properties are confounded in most nonlinear observables. We
therefore test each star against two nested nulls. Fourier (FT)
surrogates~\cite{Theiler1992} preserve only the power spectrum and
randomize the phases, so they also alter the amplitude marginal;
rejecting this null indicates departure from a linear Gaussian process
but does not localize the cause. Iterative amplitude-adjusted Fourier
(IAAFT) surrogates~\cite{Schreiber1996}, iterated for $200$ steps, preserve
\emph{both} the power spectrum and the amplitude marginal, matching the
amplitude marginal exactly and the power spectrum to a small residual, and
thereby isolate higher-order (phase) structure. A departure that survives the
FT null but not the IAAFT null is explained by the amplitude distribution
alone; a departure from the IAAFT null would instead indicate higher-order
temporal structure beyond the power spectrum and amplitude marginal. Such
higher-order structure is a necessary signature of, but is not by itself
sufficient to establish, the singular-continuous organization of
strange-nonchaotic dynamics~\cite{Pikovsky1995,Prasad2001}.

Table~\ref{tab:kepler} reports both scores for all seven targets. One
golden star, KIC~4064484, is the single most extreme member of its
Fourier ensemble ($p_\text{FT}=0.010$, rank $1/100$). Against the IAAFT
null the same star is unremarkable ($p_\text{IAAFT}=0.52$, rank
$52/100$; Fig.~\ref{fig:kepler}), and no target in the sample rejects the IAAFT null.
Because seven targets are screened, the smallest Fourier value
($p_\text{FT}=0.010$) is not family-wise significant: a Bonferroni correction over
the seven tests gives $p\approx0.07$, so the screen yields a ranked candidate
rather than a multiple-comparison-corrected detection. The
Fourier-null departure of KIC~4064484 is therefore attributable to its
non-Gaussian amplitude distribution rather than to phase structure. The
golden star KIC~9453114 does not reach the Fourier threshold under the
exchangeable statistic ($p_\text{FT}=0.060$). The Fourier ranking
does not track the golden/non-golden taxonomy either: the second most
extreme Fourier score belongs to a non-golden control, KIC~4484128
($p_\text{FT}=0.050$), consistent with the departures reflecting the
amplitude distribution rather than the pulsation structure that defines
the golden class.

\begin{table}[t]
\centering
\caption{AAH--QRC surrogate screening of the Kepler RR Lyrae light
curves against two nested nulls. Each star is scored against $K=99$
surrogates with the symmetric leave-one-out rank test; $p_\text{FT}$
uses Fourier surrogates (power spectrum only) and $p_\text{IAAFT}$ uses
IAAFT surrogates (power spectrum and amplitude marginal). The minimum
attainable value is $1/100=0.010$. No star rejects the stronger IAAFT
null.}
\label{tab:kepler}
\begin{tabular}{l l r r r r}
\hline\hline
Target & Group & $z_\text{FT}$ & $p_\text{FT}$ & $z_\text{IAAFT}$ & $p_\text{IAAFT}$ \\
\hline
KIC~4064484           & golden     & $+5.11$ & $0.010$ & $-0.22$ & $0.52$ \\
KIC~9453114           & golden     & $+2.14$ & $0.060$ & $-0.81$ & $0.78$ \\
KIC~5520878           & golden     & $+0.25$ & $0.360$ & $-0.66$ & $0.72$ \\
KIC~8832417           & golden     & $+1.57$ & $0.110$ & $+0.37$ & $0.34$ \\
KIC~4484128           & non-golden & $+2.11$ & $0.050$ & $-0.19$ & $0.51$ \\
KIC~7505345           & non-golden & $+0.82$ & $0.180$ & $-1.35$ & $0.98$ \\
KIC~5520878 surrogate & null       & $+0.01$ & $0.510$ & $-0.86$ & $0.82$ \\
\hline\hline
\end{tabular}
\end{table}

\begin{figure}[htb]
  \centering
  \includegraphics[width=0.62\textwidth]{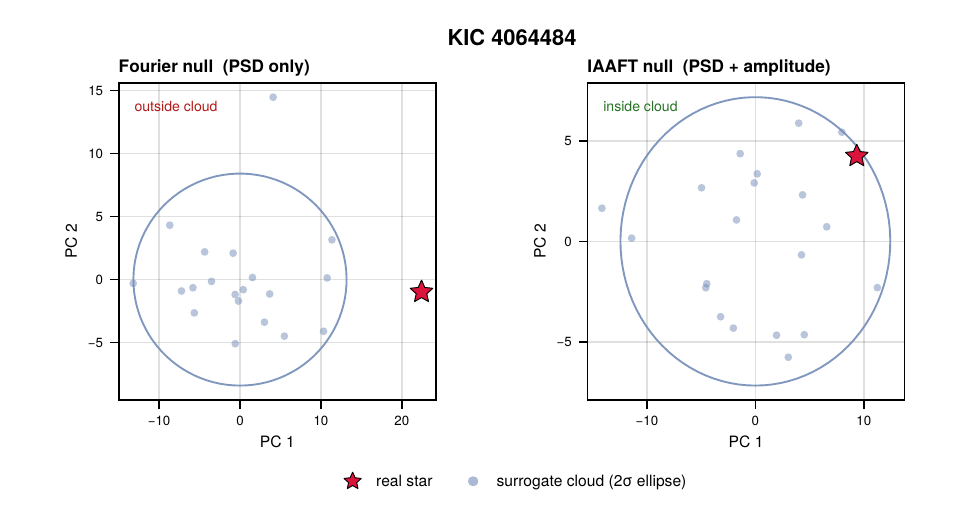}
  \caption{Screening the Kepler RR Lyrae star KIC~4064484 against two surrogate nulls.
  The reservoir feature vector ($\star$) is projected onto the leading two principal components of
  its surrogate cloud (points; $2\sigma$ ellipse), for a Fourier null (left; preserves the power
  spectrum) and the stronger IAAFT null (right; also preserves the amplitude distribution).
  The star falls outside the Fourier cloud ($p_\text{FT}=0.010$) but inside the IAAFT cloud
  ($p_\text{IAAFT}=0.52$), so the Fourier-null departure reflects the signal's amplitude distribution
  rather than higher-order phase structure. Significances use a symmetric leave-one-out rank test on
  $99$ surrogates; clouds are drawn for a $20$-surrogate subset for legibility.}
  \label{fig:kepler}
\end{figure}

The IAAFT non-detections do not contradict the strange-nonchaotic
interpretation of Lindner~\textit{et al.}~\cite{Lindner2015}. Their
argument rests on a spectral signature: the two principal pulsation
frequencies of the golden stars satisfy $f_2/f_1\approx\varphi^{-1}$,
and the multifractal analysis is performed on the power spectrum. That
signature is a property of the spectrum, which the IAAFT surrogates
preserve by construction, so it is present in every surrogate as well as
in the star. Our IAAFT test therefore does not probe the golden-ratio
structure itself; it asks whether the reservoir fingerprint carries
\emph{additional} higher-order phase organization beyond the spectrum
and amplitude marginal. A star inside the IAAFT cloud simply has no such
excess structure detectable at this record length, which is orthogonal
to, rather than in tension with, a spectral SNA diagnosis. The two
approaches probe complementary aspects of the same signal.

KIC~4064484 departs from the linear Gaussian null but not from the
amplitude-preserving null. The present analysis therefore ranks it as a
candidate for further study but does not establish strange-nonchaotic phase
structure in any of the stars. Confirmatory inference would require longer
records, gap-aware preprocessing, or supervised calibration.


\begin{thebibliography}{40}

\bibitem{Fujii2017}
K.~Fujii and K.~Nakajima,
Harnessing disordered-ensemble quantum dynamics for machine learning,
Phys.\ Rev.\ Applied \textbf{8}, 024030 (2017).

\bibitem{Nakajima2019}
K.~Nakajima and I.~Fischer, eds.,
\emph{Reservoir Computing: Theory, Physical Implementations, and Applications}
(Springer, Singapore, 2021).

\bibitem{Mujal2021}
P.~Mujal, R.~Mart\'{\i}nez-Pe\~{n}a, J.~Nokkala, J.~Garc\'{\i}a-Beni,
G.~L.\ Giorgi, M.~C.\ Soriano, and R.~Zambrini,
Opportunities in quantum reservoir computing and extreme learning machines,
Adv.\ Quantum Technol.\ \textbf{4}, 2100027 (2021).

\bibitem{Beer2020}
K.~Beer, D.~Bondarenko, T.~Farrelly, T.~J.\ Osborne, R.~Salzmann, D.~Scheiermann,
and R.~Wolf,
Training deep quantum neural networks,
Nat.\ Commun.\ \textbf{11}, 808 (2020).

\bibitem{Ghosh2019}
S.~Ghosh, A.~Opala, M.~Matuszewski, T.~Paterek, and T.~C.\ H.\ Liew,
Quantum reservoir processing,
npj Quantum Inf.\ \textbf{5}, 35 (2019).

\bibitem{Nokkala2021}
J.~Nokkala, R.~Mart\'{\i}nez-Pe\~{n}a, G.~L.\ Giorgi, V.~Parigi,
M.~C.\ Soriano, and R.~Zambrini,
Gaussian states of continuous-variable quantum systems provide universal and
versatile reservoir computing,
Commun.\ Phys.\ \textbf{4}, 53 (2021).

\bibitem{Martinez2021}
R.~Mart\'{\i}nez-Pe\~{n}a, G.~L.\ Giorgi, J.~Nokkala, M.~C.\ Soriano,
and R.~Zambrini,
Dynamical phase transitions in quantum reservoir computing,
Phys.\ Rev.\ Lett.\ \textbf{127}, 100502 (2021).

\bibitem{Xia2022}
W.~Xia, J.~Zou, X.~Qiu, F.~Chen, B.~Zhu, C.~Li, D.-L.~Deng, and X.~Li,
Configured quantum reservoir computing for multi-task machine learning,
Sci.\ Bull.\ \textbf{68}, 2321 (2023).

\bibitem{Grebogi1984}
C.~Grebogi, E.~Ott, S.~Pelikan, and J.~A.\ Yorke,
Strange attractors that are not chaotic,
Physica D \textbf{13}, 261 (1984).

\bibitem{Feudel2006}
U.~Feudel, S.~Kuznetsov, and A.~Pikovsky,
\emph{Strange Nonchaotic Attractors: Dynamics between Order and Chaos in
Quasiperiodically Forced Systems}
(World Scientific, Singapore, 2006).

\bibitem{Prasad2001}
A.~Prasad, S.~S.\ Negi, and R.~Ramaswamy,
Strange nonchaotic attractors,
Int.\ J.\ Bifurc.\ Chaos \textbf{11}, 291 (2001).

\bibitem{Romeiras1987}
F.~J.\ Romeiras and E.~Ott,
Strange nonchaotic attractors of the damped pendulum with quasiperiodic forcing,
Phys.\ Rev.\ A \textbf{35}, 4404 (1987).

\bibitem{Pikovsky1995}
A.~S.\ Pikovsky and U.~Feudel,
Characterizing strange nonchaotic attractors,
Chaos \textbf{5}, 253 (1995).

\bibitem{HeagyHammel1994}
J.~F.~Heagy and S.~M.~Hammel,
The birth of strange nonchaotic attractors,
Physica D \textbf{70}, 140 (1994).

\bibitem{Aubry1980}
S.~Aubry and G.~Andr\'{e},
Analyticity breaking and Anderson localization in incommensurate lattices,
Ann.\ Israel Phys.\ Soc.\ \textbf{3}, 133 (1980).

\bibitem{Harper1955}
P.~G.\ Harper,
Single band motion of conduction electrons in a uniform magnetic field,
Proc.\ Phys.\ Soc.\ A \textbf{68}, 874 (1955).

\bibitem{Avila2009}
A.~Avila and S.~Jitomirskaya,
The ten Martini problem,
Ann.\ Math.\ \textbf{170}, 303 (2009).

\bibitem{Bellissard1982}
J.~Bellissard, R.~Lima, and D.~Testard,
A metal-insulator transition for the almost Mathieu model,
Commun.\ Math.\ Phys.\ \textbf{88}, 207 (1983).

\bibitem{Lindblad1976}
G.~Lindblad,
On the generators of quantum dynamical semigroups,
Commun.\ Math.\ Phys.\ \textbf{48}, 119 (1976).

\bibitem{GKS1976}
V.~Gorini, A.~Kossakowski, and E.~C.~G.~Sudarshan,
Completely positive dynamical semigroups of $N$-level systems,
J.\ Math.\ Phys.\ \textbf{17}, 821 (1976).

\bibitem{Roushan2017}
P.~Roushan \emph{et al.},
Spectroscopic signatures of localization with interacting photons in
superconducting qubits,
Science \textbf{358}, 1175 (2017).

\bibitem{Prasad1997}
A.~Prasad, V.~Mehra, and R.~Ramaswamy,
Strange nonchaotic attractors in the quasiperiodically forced logistic map,
Phys.\ Rev.\ E \textbf{57}, 1576 (1998).

\bibitem{Ding1989}
M.~Ding, C.~Grebogi, and E.~Ott,
Dimensions of strange nonchaotic attractors,
Phys.\ Lett.\ A \textbf{137}, 167 (1989).

\bibitem{Jaeger2002}
H.~Jaeger,
Short term memory in echo state networks,
GMD Report~152 (2002).

\bibitem{Appeltant2011}
L.~Appeltant, M.~C.~Soriano, G.~Van der Sande, J.~Danckaert, S.~Massar,
J.~Dambre, B.~Schrauwen, C.~R.~Mirasso, and I.~Fischer,
Information processing using a single dynamical node as complex system,
Nat.\ Commun.\ \textbf{2}, 468 (2011).

\bibitem{Larger2012}
L.~Larger, M.~C.~Soriano, D.~Brunner, L.~Appeltant, J.~M.~Guti\'{e}rrez,
L.~Pesquera, C.~R.~Mirasso, and I.~Fischer,
Photonic information processing beyond Turing: an optoelectronic
implementation of reservoir computing,
Opt.\ Express \textbf{20}, 3241 (2012).

\bibitem{Anderson1958}
P.~W.~Anderson,
Absence of diffusion in certain random lattices,
Phys.\ Rev.\ \textbf{109}, 1492 (1958).

\bibitem{Evers2008}
F.~Evers and A.~D.\ Mirlin,
Anderson transitions,
Rev.\ Mod.\ Phys.\ \textbf{80}, 1355 (2008).

\bibitem{Lindner2015}
J.~F.\ Lindner, V.~Kohar, B.~Kia, M.~Hippke, J.~G.\ Learned, and
W.~L.\ Ditto, Strange nonchaotic stars,
Phys.\ Rev.\ Lett.\ \textbf{114}, 054101 (2015).

\bibitem{Theiler1992}
J.~Theiler, S.~Eubank, A.~Longtin, B.~Galdrikian, and J.~D.\ Farmer,
Testing for nonlinearity in time series: the method of surrogate data,
Physica D \textbf{58}, 77 (1992).

\bibitem{Schreiber1996}
T.~Schreiber and A.~Schmitz,
Improved surrogate data for nonlinearity tests,
Phys.\ Rev.\ Lett.\ \textbf{77}, 635 (1996).

\bibitem{suppmat}
See Supplemental Material at [URL will be inserted by publisher] for full
derivations, signal generation details, and additional figures.

\end{thebibliography}
\end{document}